\documentclass{agujournal}
\draftfalse

\journalname{Journal of Advances in Modeling Earth Systems}

\usepackage{booktabs}
\usepackage{multirow}
\usepackage{mathtools}
\usepackage[hidelinks]{hyperref}

\begin{document}

\title{Coastal Tropical Convection in a Stochastic Modeling Framework}

%
%

\authors{Martin Bergemann\affil{1,3}, Boualem Khouider\affil{2}, Christian Jakob\affil{1,3}}

\affiliation{1}{School of Earth, Atmosphere and Environment, Faculty of Science, Monash University, Melbourne, VIC 3000, Australia}
\affiliation{2}{Department of Mathematics and Statistics, University of Victoria, Victoria, B.C. Canada V8W 3P4}
\affiliation{3}{ARC Centre of Excellence for Climate System Science}
\correspondingauthor{Martin Bergemann}{martin.bergemann@monash.edu}



\begin{keypoints}
\item Coastal convection exhibits characteristic behavior that is poorly captured in global climate models. 
\item A trigger function is developed to decide if convection is supported by coastal effects.
\item The application of this trigger function can improve the representation of tropical convection near coasts considerably.
\end{keypoints}

%

\begin{abstract}
Recent research has suggested that the overall dependence of convection near coasts on large-scale atmospheric conditions is weaker than over the open ocean or inland areas. This is due to the fact that in coastal regions convection is often supported by meso-scale land-sea interactions and the topography of coastal areas. As these effects are not resolved and not included in standard cumulus parametrization schemes, coastal convection is among the most poorly simulated phenomena in global models. To outline a possible parametrization framework for coastal convection we develop an idealized modeling approach and test its ability to capture the main characteristics  of coastal convection. The new approach first develops a decision algorithm, or trigger function, for the existence of coastal convection. The function is then applied in a stochastic cloud model to increase the occurrence probability of deep convection when land-sea interactions are diagnosed to be important. The results suggest that the combination of the trigger function with a stochastic model is able to capture the occurrence of deep convection in atmospheric conditions often found for coastal convection. When coastal effects are deemed to be present the spatial and temporal organization of clouds that has been documented form observations is well captured by the model. The presented modeling approach has therefore potential to improve the representation of clouds and convection in global numerical weather forecasting and climate models.
\end{abstract}

%
%

\section{Introduction}
In coastal areas of the tropics precipitation variance is strongly influenced by diurnal and sub-diurnal frequencies \citep{Yang2001}. The diurnal cycle of precipitation is therefore a prominent mode in convective systems that are associated with tropical rainfall. Hence observations of the diurnal precipitation cycle in coastal areas of the tropics have often been a subject of intensive studies \citep[][a.o.]{Kousky1980,Geotis1985,Skinner1994,Oliphant2001,Mapes2003a,Kondo2006,Zhuo2013,Peatman2014}. 
The vast majority of theses studies utilized spaceborne observations to identify and understand the key-mechanisms of the structure and behavior of the diurnal cycle of rainfall, clouds and convective systems. Studies using satellite rainfall estimates and focusing on the Maritime Continent showed that rainfall in this area exhibit characteristic patterns \citep{Williams1987,Ohsawa2001}.  These patterns are often organized by coastlines \citep{Holland1980} with rain between 2100 LT and 0900 LT concentrated over the oceans peaking in the early morning while the 0900 LT to 2100 LT precipitation is mainly located over land with maxima occurring in the early evening. It has been argued that one of the important mechanisms that cause this characteristic spatio- temporal organization are land-sea breeze effects \citep{Mori2004,Qian2008}. Using an objective pattern recognition algorithm, \citet{Bergemann2015} showed that in coastal areas of the tropics approximately one third of the total rainfall amount is associated with these meso-scale circulation systems. The details of any land-sea breeze circulation are dependent on multiple factors, where the most prominent ones are coastal arrangement, orography, and variations due to the Coriolis effect \citep{Haurwitz1947,Rotunno1983}. 

In coastal regions of the tropics, such as the Maritime Continent, global numerical weather prediction and climate models show large errors in rainfall \citep{Yang2001,Neale2002,Nguyen2015}. Here rainfall is usually underestimated over land and overestimated over the ocean, which indicates that the complex processes associated with coastal land-sea interaction are poorly captured \citep[e.g][]{Mapes2003b,Slingo2004,Gianotti2011}. \citet{Hohenegger2015} investigated the coupling between convection and sea-breeze characteristics at different model resolutions and found that not only the presence of sea-breeze can be influential for convection also the models representation of convection can have significant impact on the sea-breeze propagation.

Recent studies have shown that rainfall that is affected by meso-scale land-sea interactions is significantly less dependent on large-scale atmospheric conditions \citep{Birch2016,Bergemann2016}. This finding is of relevance to the modeling community because the topography that influences coastal convection is not fully resolved in global numerical models and hence tropical coastal convection remains poorly captured by them. 

In global numerical weather prediction and climate models many important processes, like coastal land-sea interactions, are not fully or not at all resolved. Designing models that translate the resolved scales of the climate model into the unresolved processes and providing feedback from the unresolved to resolved scales are the two key task of parametrizations \citep{Arakawa2004}. Usually numerical weather prediction and climate models apply deterministic parametrizations of theses unresolved processes. Theses models are usually very idealized  and conceptual \citep[e.g][]{Arakawa1974,Tiedtke1989,Gregory1990}  and hence neglecting various processes that are known to be important for the presence and amplification of moist convection. \citet{Bergemann2016} argued that "it is necessary to enhance current cumulus parametrizations to be able to model tropical rainfall associated with coastlines". The aim of this study is therefore to propose a computationally simple modeling approach that is able to capture the key characteristics of coastal convection which have been discussed by previous studies \citep[e.g][]{Holland1980,Mori2004,Bergemann2015,Bergemann2016}. This study addresses this issue by developing a method to identify potential sea-breeze conditions solely based on large-scale atmospheric conditions. The method will be then applied in the \emph{stochastic multi cloud model} \cite[SMCM,][]{Khouider2010} to test its ability to represent some of the key characteristics of convection in coastal areas. We chose the SMCM because despite it's simplicity it has been successfully applied in various General Circulation models. For instance coupled \citet{Peters2017} the SMCM to the state of the art general circulation model ECHAM6 and demonstrated that the representation of the Madden-Julian-Oscillation [MJO] is improved when compared to observations. Similarly, \citet{Goswami2017} successfully applied the SMCM in NCEP's Climate Forecasting model. \citet{Deng2015,Deng2016} and \citet{Ajayamohan2016} utilized the model as a cumulus parametrization in an aquaplanet GCM to simulate the  and monsoon-like intra-seasonal oscillations. Because the SMCM has been confirmed to enhance the representation of tropical convection it has potential to help to improve the simulation of coastal convection. 

This study is divided into three parts. The first part introduces the method that identifies conditions favorable for sea-breeze conditions [Section \ref{sec:trigger}]. In Section \ref{sec:smcm} the modeling approach to mimic coastal clouds is introduced. The main result, presented in Section \ref{sec:results}, demonstrates the capability of the new model to capture some of the key features of convection in the coastal tropics. This is followed by a summary and conclusion in Section \ref{sec:con}.
 
\section{A a new Trigger Function for Coastal Convection}
\label{sec:trigger}
\subsection{Overview}
The first step in modeling coastally affected convection is to decide whether or not the convection at a specific location and time is influenced by coastal effects. One commonly used approach in parametrizing atmospheric phenomena, especially those associated with tropical convection, is the use of a trigger function \citep{Suhas2014,Hottovy2015}. Any function that describes coastal effects such as land-sea breeze convergence would take the shape of the coastline into account. When the coastline is irregular, local regions of enhanced or weakened low-level convergence may develop \citep{McPherson1970}. 
The evolution of the land-sea breeze is more complicated when the prevailing synoptic wind-regime is taken into account \citep{Jiang2012}.
\begin{figure}[t]
\centering
\includegraphics[width=\textwidth]{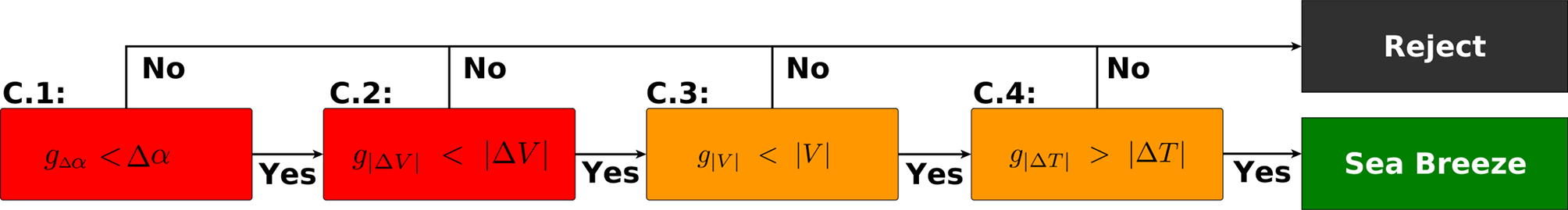}
\caption{The sea-breeze filter with the applied thresholds for wind direction [$\Delta \alpha$], magnitude of wind speed [$|V|$], the change of wind speed magnitude [$|\Delta V|$] and thermal heating contrast between land and ocean [$\Delta T=T_{\mathrm{land}} - T_{\mathrm{ocean}}$] are visualized in this flow chart. $g_x$ are the time series of the quantities that are considered.}
\label{fig:5-03}
\end{figure}
The interaction of local wind systems associated with land-sea breeze convergence and synoptic wind patterns can become very complicated for more complex coastlines such as those found over the Maritime Continent. A simpler method describing the strength of the land-sea breeze, independent of the shape of the coastline is highly desirable, especially in global models that do not capture the full complexity of most coastlines. \citet{Borne1998} developed a filtering method, that is independent of the shape of the coastline and objectively identifies potential sea-breeze days.
The simplicity of this approach makes it a good candidate for a trigger function of coastally associated land-sea interaction. The filtering technique takes synoptic lower-tropospheric and surface wind conditions as well as thermal heating contrast between land and ocean into account and considers the following six conditions:
\begin{enumerate}
\item[C.1] It is assumed that land-sea breeze systems can only develop under synoptically stable conditions. Therefore the first condition considers the change in wind direction at 700 hPa level within 24 hours. Here it is assumed that the local land-sea-breeze can only propagate on- or offshore if the change in large-scale wind direction [$g_{\Delta \alpha}$] within 24 hours is less then or equal to 90\textdegree.
\item[C.2] In addition to the change of wind direction $g_{\Delta \alpha}$, the change of its magnitude $g_{|\Delta V|}$ has to be taken into account. A threshold of \mbox{$|\Delta V|=$ 6 $\frac{\mathrm{m}}{\mathrm{s}}$} for the maximum \emph{change} of synoptic wind speed $|V|$ within the last 24 hours is applied in this condition.
\item[C.3] With the help of a nonlinear numerical model \citet{Arritt1993} showed that sea-breeze convergence can only exist in large-scale ambient wind-flows of up to \mbox{$|V|$= 11 $\frac{\mathrm{m}}{\mathrm{s}}$}. This wind speed was chosen by \citet{Borne1998} as the maximum synoptic wind speed for sea-breeze conditions.
\item[C.4] An important criteria for the development of sea-breeze conditions is the thermal heating contrast between land and ocean $g_{|\Delta T|}$. It is assumed that the mean thermal heating contrast over a time period of 24 h should be greater or equal \mbox{3 K}.
\item[C.5] During the build up of the sea breeze the surface winds should at least change by 30\textdegree. The build up period for the sea breeze is considered to be the time from sunrise +1 h to sunset -5 h.
\item[C.6] The last condition considers the surface winds after an abrupt sea-breeze induced change of surface winds during sunrise +1 to sunset. 
\cite{Borne1998} assume that if the sea-breeze induced surface wind change [$g_{|\delta V_{s}|}$] occurs at time $t_0$ then the mean wind change $\overline{g}_{|\delta V_{s}|}$ between $t_0$ and $t_{+5\mathrm{h}}$ is $\leq \frac{1}{6} \cdot g_{|\delta V_{s}|}$.
\end{enumerate}
The application of conditions \emph{C.1} to \emph{C.4} is straight forward in a global modeling framework because only large-scale properties are considered. Conditions \emph{C.5} and \emph{C.6} on the other hand are problematic because they make use of surface wind properties which in a model are the result of a boundary-layer parametrization that doesn't take coastal effects into account. Therefore the last two conditions \emph{C.5} and \emph{C.6} are neglected for building the trigger function for coastal processes. The decision process adopted here is visualized in Figure \ref{fig:5-03}.
\subsection{Data}
\begin{figure}[t]
\centering
\includegraphics[width=\textwidth]{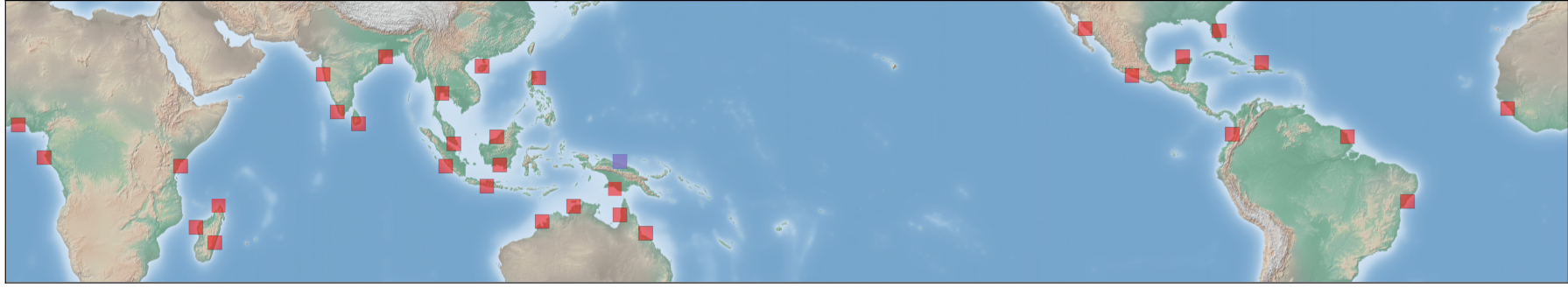}
\caption{Location of the 33 $150 \times 150$ km$^2$ locations. For each box the sea-breeze filter is forced with the area averaged atmospheric conditions from the ERA-I reanalysis data.}
\label{fig:5-04}
\end{figure}

The coastal effect trigger function should identify potential sea-breeze conditions for an area of a size that is roughly that of a climate model grid box. Therefore 33 different coastal locations with a spatial scale of roughly $150 \times 150$ km$^2$ in the tropics are chosen and the above described filtering technique is applied in each of them using area averaged atmospheric conditions in each box [Figure \ref{fig:5-04}]. The input data - large-scale wind $g_{|\vec{V}|}$ at pressure level $p$ and thermal heating contrast $g_{\Delta\mathrm{T}}$ - are calculated from the ERA-Interim reanalysis [ERA-I] with a spatial resolution of $0.75$\textdegree{} every 6 hours in time. The considered period is 1998 to 2016. The locations were chosen that half of the box [2 ERA-I grid points] is land and the other half ocean. The thermal heating contrast between land and ocean is defined as $\Delta\mathrm{T} = \Theta_l\cdot\mathrm{\overline{T}} - \Theta_s \cdot \mathrm{\overline{T}}$. With $\Theta_{l,s}$ the land and sea points in the ERA-I land-sea mask that corresponds to the chosen location.
\subsection{Estimation of the Input Thresholds}
\citet{Borne1998} developed the method for the middle latitudes and  neither tested its application in tropical areas nor did they evaluate the sensitivity of the applied thresholds. In the tropics the large-scale pressure and temperature fields are usually much more homogeneous than in the middle and high latitudes and it is unclear if the limits \citet{Borne1998} were using can be applied to tropical synoptic conditions. We wish to find a threshold setup that captures sea breeze conditions in the tropics as much as possible while the false alarm rate remains low. We are not aware of any existing global dataset that describes the presence of sea-breeze conditions over land and ocean. Yet, \citet{Bergemann2015} presented for the first time an algorithm that objectively finds precipitation features in spaceborne rainfall estimates that can be associated with coastal land-sea interaction. By definition this dataset is closely related to sea-breeze conditions in the tropics; when coastally affected rainfall occurs sea-breeze conditions should also be present. Sea-breeze conditions alone do not guarantee the occurrence of coastal rainfall. To find a reasonable threshold setup for the coastal trigger function we correlate the occurrence of coastal rain [$>0$ mm/3h] with the trigger function and find the maximum of the correlation. The coastal rainfall data is based on CMORPH satellite based rainfall estimates \citep{Joyce2004}. The dataset has a spatial resolution of 0.25\textdegree and 3 hours in time. The considered time period is 1998 to 2016.
\begin{figure}
\includegraphics[width=\textwidth]{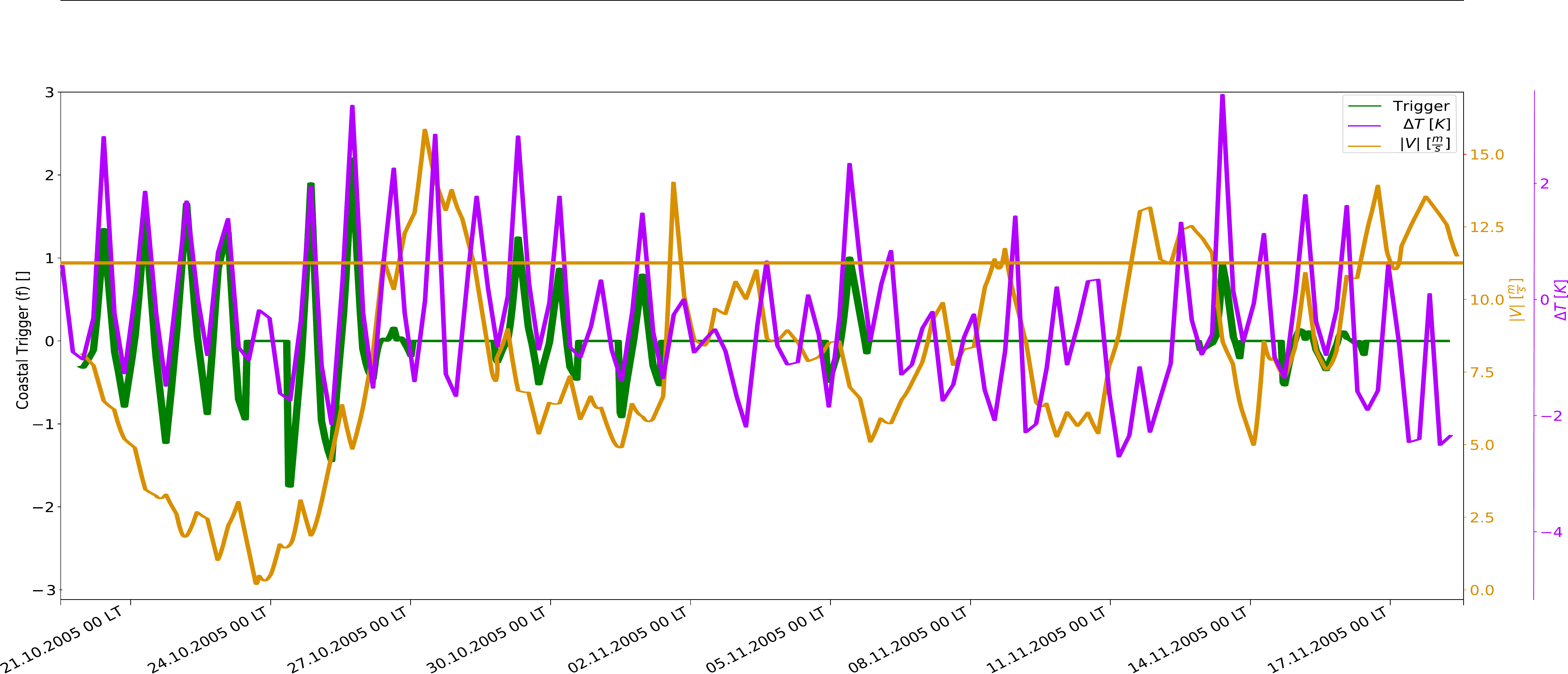}
\caption{Example of a one month time series of the coastal trigger function [green] the thermal heating contrast between land and ocean $\Delta T$ [purple] and the wind speed at 700 hPa $|\vec{V}|$ over Darwin, Austrialia. The solid horizontal lines indicates a wind speed threshold of 11 $\frac{\mathrm{m}}{\mathrm{s}}$.}
\label{fig:5-02-1}
\end{figure}
Finding the optimal correlation with four independent trigger input variables is computationally expensive and it is very hard to find a global maximum. To identify the most important variables and reduce the dimension of the optimization we first apply a variance based sensitivity analysis after \citet{Saltelli2008}. Here we create an ensemble of input thresholds and the number of sea-breeze days each ensemble member. The sea-breeze filter ensemble variance is then decomposed into fractions of variances that are related to each input variable. A more detailed description of the procedure is given in Appendix \ref{sec:appendix}. We also consider the large-scale wind conditions at all levels between 700 - 850 hPa in 50 hPa increments and vary the time period considered for the change of wind speed and direction from 24 to 12 hours. Table \ref{tab:5-01} summarizes the results of the variance based sensitivity test for all 33 tropical coastal locations.

\begin{table}
\caption{a) Fractions of variance [in \%] for each input threshold that contribute to the total sea-breeze filter variance. The columns represent the application of the filter on different pressure levels. a) The time period considered for changing conditions is 24 hours. b) As a) but for a time period of 12 hours.}
\begin{tabular}{l r c c c c |r}\toprule
& &700 hPa & 750 hPa & 800 hPa & \multicolumn{1}{c}{850 hPa} \\ \midrule
a)&$\Delta T$ & $71.7 \pm 24.6$ & $70.4 \pm 21.9$ & $65.7 \pm 15.1$ &$66.1 \pm 17.0$ &\multirow{4}{*}{24 h} \\
&$|\vec{V}|$ & $6.04 \pm 7.12$ & $7.19 \pm 6.21$ & $8.53 \pm 5.17$ & $8.04 \pm 7.82$ \\
&$\Delta|\vec{V}|$ & $0.59 \pm 0.95$ & $0.69 \pm 1.05$ & $0.97 \pm 0.92$ & $1.01 \pm 1.00$\\
&$\Delta \alpha$ & $0.14 \pm 0.21$ & $0.07 \pm 0.13$ & $0.24 \pm 0.17$ & $0.24 \pm 0.18$\\ \midrule
b)&$\Delta T$ & $68.4 \pm 23.2$ & $69.3 \pm 19.9$ & $67.9 \pm 13.5$ &$69.2 \pm 19.2$ &\multirow{4}{*}{12 h} \\
&$|\vec{V}|$ & $7.06 \pm 6.35$ & $7.89 \pm 5.29$ & $7.98 \pm 5.65$ & $8.94 \pm 8.33$ \\
&$\Delta|\vec{V}|$ & $0.59 \pm 1.05$ & $0.66 \pm 0.89$ & $0.93 \pm 0.86$ & $1.10 \pm 0.91$\\
&$\Delta \alpha$ & $0.21 \pm 0.17$ & $0.17 \pm 0.19$ & $0.34 \pm 0.12$ & $0.58 \pm 0.36$\\
\bottomrule
\end{tabular}
\label{tab:5-01}
\end{table}
The sensitivity test shows that there is little impact on the choice of pressure level and time period. It can also be seen that most of variance is attributed to the thermal heating contrast between land and ocean. The changes of wind speed and direction contribute only very little to the total variance of the filtering method. This can be explained by the relatively steady spatial and temporal conditions in the tropics. We therefore set thresholds for these two variables to be identical to the ones chosen by \citet{Borne1998}. Figure \ref{fig:5-02-1} serves as an example to visualize the impact of the thermal heating contrast $\Delta T$ and the synoptic wind speed $|\vec{V}|$. During times when the wind speed is below the applied threshold [in the present case $11$ m/s] the trigger function is mainly modulated by variations in thermal heating contrast. 

\begin{figure}
\centering
\includegraphics[width=\textwidth]{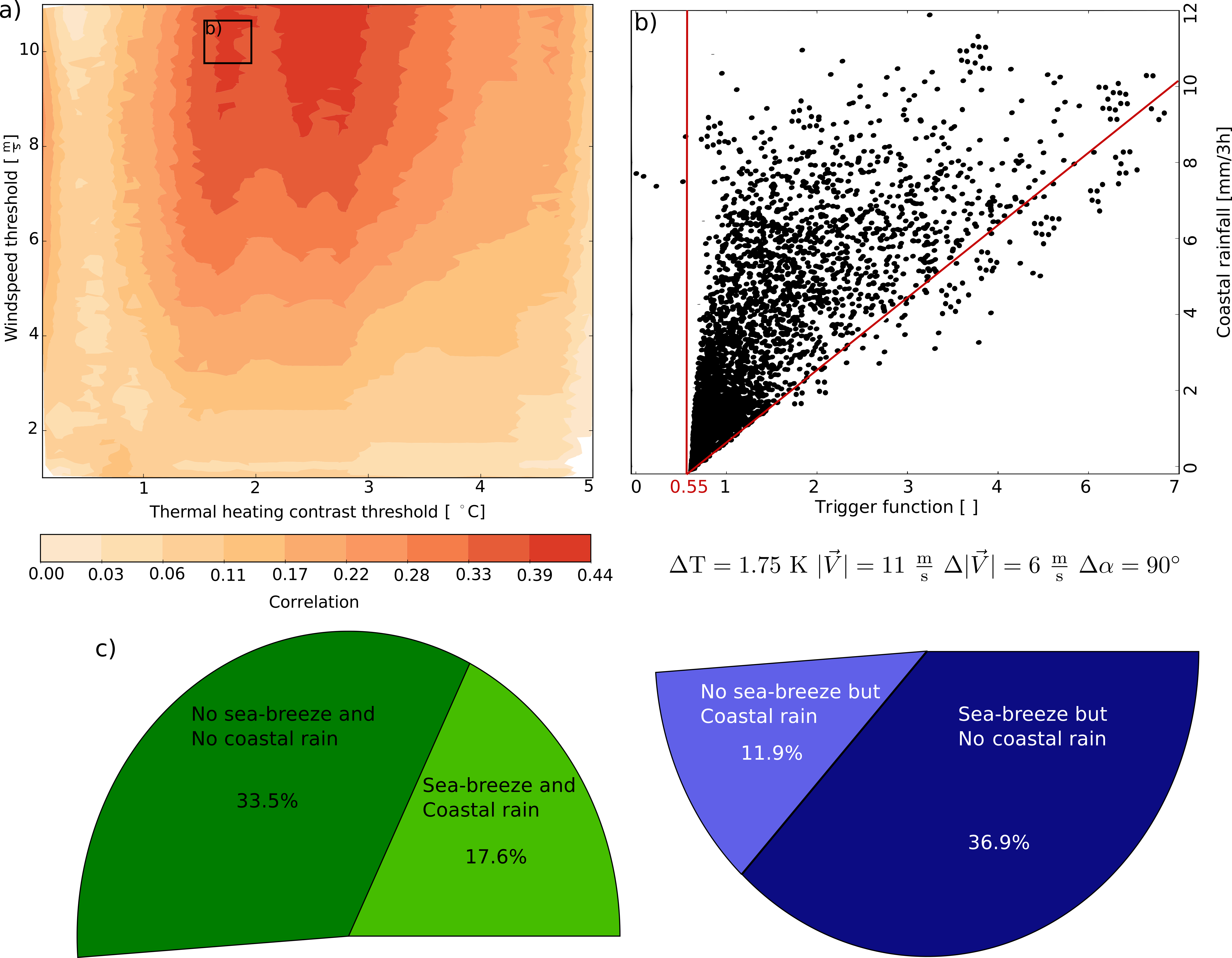} 
\caption{a) The correlation of coastally affected rainfall data after \citet{Bergemann2015} with the scaled version of the sea-breeze trigger as a function of thresholds applied to thermal heating contrast and wind-speed. b) The scaled sea-breeze trigger is plotted against the coastal rainfall for the thermal heating contrast of 1.75 K and a wind speed threshold of 11 $\frac{\mathrm{m}}{\mathrm{s}}$ [this threshold setup is marked in the box in a]. c) Percentages of agreement of the filter function and the coastal rainfall [green] and no agreement of the two datasets [red].}
\label{fig:5-05}
\end{figure}

Both, the variance decomposition and Figure \ref{fig:5-02-1} show that the threshold choices of maximum wind speed and especially thermal heating contrast can have large impacts on the results of the sea-breeze filtering method and their thresholds must be carefully chosen. To find a reasonable threshold setup we now try to optimize the two values using the data set for coastally influenced rainfall \citep{Bergemann2015}. The filtering method described above finds only conditions that are likely to have a land-sea breeze convergence and applies yes/no decisions but no information about the strength of the land-sea breeze circulation is derived in the process. An optimization can only be applied if the output values of the filter are continuous. The simplest method to make the binary output continuous is scaling the output by thermal heating contrast and wind speed. Thermal heating contrast and wind speed are chosen because the two variables are most influential to the outcome of the filtering process. If $f(t)$ is the binary output of the sea-breeze filter method and ${\Delta\mathrm{T}}$ and ${|\vec{V}|}$ are the thresholds that are applied then the following scaling relationships can be implemented to produce continuous results:
\begin{linenomath*}
\begin{equation} \label{eq:5-6}
\tilde{f}(t) = \left\lbrace \begin{matrix}
0 & \mathrm{if\ } f(t) = 0 & \ \\
\underbrace{\frac{|g_{\Delta\mathrm{T}}(t)|}{{\Delta\mathrm{T}}}}_{>0} \cdot \underbrace{\frac{{|\vec{V}|}-g_{|\vec{V}|}(t)}{{|\vec{V}|}}}_{>0}  & \mathrm{if\ } f(t) = 1& \ \\
\end{matrix} \right.
\end{equation}
\end{linenomath*}
$|g_{\Delta\mathrm{T}}(t)|$ and $g_{|\vec{V}|}(t)$ are the time series of magnitudes of thermal heating contrast and wind speed. 
\begin{table}
\caption{Setup of the input variables for the optimization process that is applied in Figure \ref{fig:5-05}-a}
\centering
\begin{tabular}{p{0.01\textwidth} c c c c c c c p{0.01\textwidth}}\toprule
& $\Delta$T & $|\vec{V}|$ & $\Delta|\vec{V}|$ & $\Delta \alpha$ & p-level & t-period& \\ \cline{2-7}
& varies & varies & 6$\frac{\mathrm{m}}{\mathrm{s}}$ & 90\textdegree & 800 hPa & 12 h& \\ \bottomrule
\end{tabular}
\label{tab:5-02}
\end{table}
Having made the output of the sea-breeze filtering process continuous an optimal threshold setup for the thermal heating contrast and the wind speed can be found by finding the maximum correlation of the $\tilde{f}$ in Equation \ref{eq:5-6} with coastally associated rainfall after \citet{Bergemann2015}. To find an optimal threshold setup we first calculate the correlation for days when coastally affected rainfall is present, on average approximately 52\% of the time. After choosing reasonable thresholds the output of the coastal trigger is then compared with the total coastal rainfall data that contains rainy and non-rainy days [Figure \ref{fig:5-05}-c]. The correlation of the scaled sea-breeze trigger $\tilde{f}$ as function of thermal heating contrast and large-scale wind speed thresholds, $|V|$ and $\Delta T$, with coastally affected rainfall is presented in Figure \ref{fig:5-05}-a. Because they have been shown to have secondary impact, the change in wind direction $\Delta \alpha$ and wind speed $\Delta |V|$ as well as the time period are chosen to be fixed in this analysis. Table \ref{tab:5-02} summarizes the parameters used in for the optimization process.

Figure \ref{fig:5-05}-a shows correlation maxima for wind speed thresholds around $\approx \ 11\frac{\mathrm{m}}{\mathrm{s}}$ and thermal heating contrasts of $\approx 1.75$K.  Figure \ref{fig:5-05}-b shows the trigger function plotted against the coastally affected rainfall for this threshold setup. The shape of the scatter shows a clear minimum value of the trigger function, at $\approx$ 0.55, below which there is almost no coastal rain. This means when coastal rain is present the trigger function is almost always greater than this value. The Figure also shows that the intensity of coastal rain scales to some extent with the magnitude of the trigger function. If the intensity of the sea-breeze convergence increases the minimum coastal precipitation also increases. This behavior indicates skill in the trigger function in determining whether or not convection is supported and enhanced by coastal effects in a modeling framework. If the above trigger function is applied for coastally associated convection one immediate question is then how much of the coastal rainfall is missed by it. Figure \ref{fig:5-05}-c shows the percentages of agreement and disagreement of the unscaled [binary] trigger function and the presence of coastal rain after \citet[][$>0$ mm/3h]{Bergemann2015}. In more than half of the cases the trigger function and coastally affected rain dataset are in agreement. Yet there remains a high percentage of cases when the filter assigns sea-breeze days without the presence of coastal precipitation. It might be tempting to label these cases as false alarm but the reader is reminded here that the applied function serves as a \emph{trigger} that can initiate convection in principle rather than a strict predictor of its presence. If the large-scale conditions are not favorable to generate deep convection then there might be a sea-breeze without the presence of rainfall. The case when the filter doesn't assign a sea-breeze day when coastal rainfall is present should be labeled as a miss. With 11.9\% the fraction of missed cases is rather small. 

The analysis shows that our method to detect sea-breeze days based solely on synoptic-scale conditions can be applied for conditions in the tropics. The scaled version of the adopted filter has potential to serve as a trigger function that can initiate convection which is associated with coastal land-sea interaction. The presented trigger function is only the first step for building a conceptual model that describes clouds in coastal tropical areas. The next Section will discuss a cloud modeling framework where this new trigger function is applied.

\section{A Cloud Model for Coastal Convection}
\label{sec:smcm}
\subsection{The Stochastic Multi Cloud Model [SMCM]}
On the resolved scale climate models describe processes  of $\approx \ O(50\ \mathrm{km})$ to $O(300\ \mathrm{km})$. Cloud processes, especially those associated with tropical convection, are usually acting on scales of $\approx \ O(100 \ \mathrm{m})$ to $O(10 \ \mathrm{km})$. The application of cloud modeling approaches in parametrizations should meet several criteria. Different cloud types with various cloud top heights distribute moisture and heat in the troposphere in different ways. For instance are stratiform clouds important for the atmosphere's radiation budget and \citet{Arakawa1974} already identified detrained condensed water from 'hot cumulus towers' as an important source of this cloudtypes. As a consequence most parametrization schemes have focused on cumulus detrainment to improve the representation of stratiform clouds \citep[e.g][]{Tiedtke1993,Randall1999}. Yet conditions controlling entrainment and detrainment of condensed water are poorly represented in traditional parametrization approaches \citep{Randall2003}. 

\begin{figure}
\centering
\includegraphics[width=\textwidth]{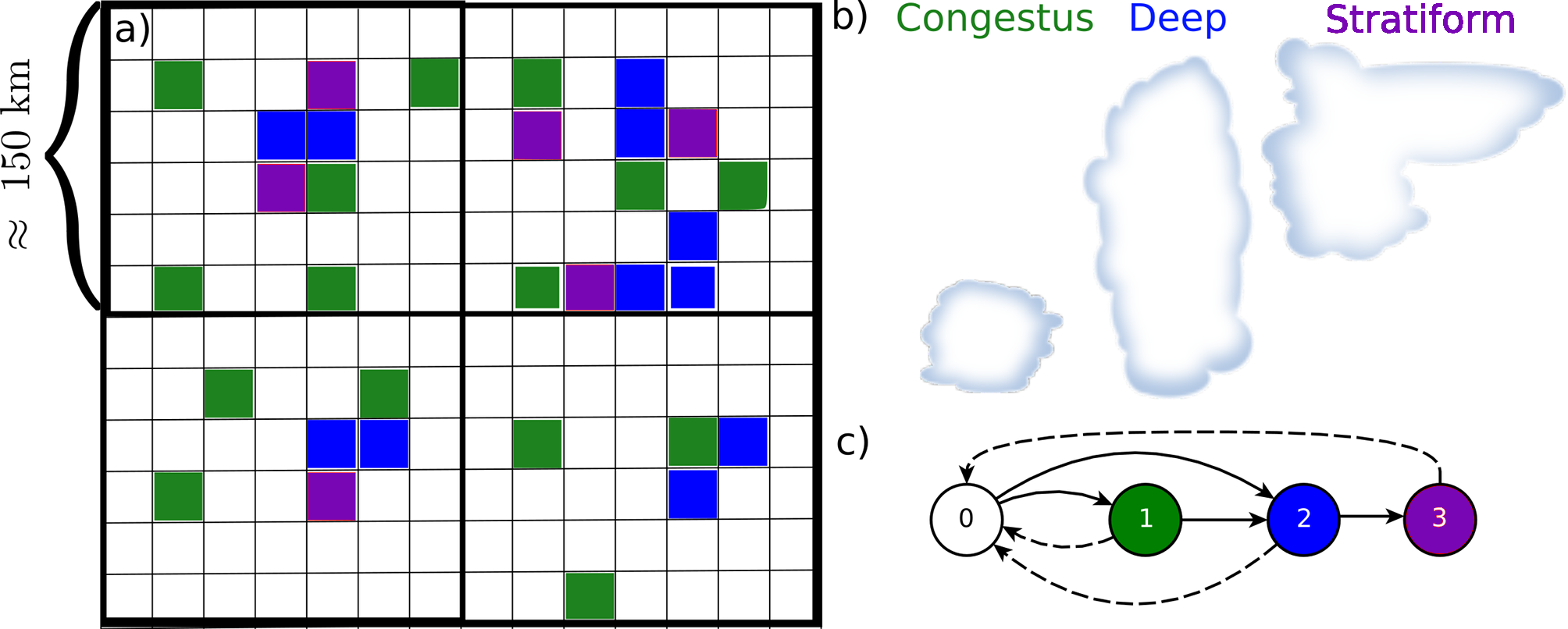}
\caption{a) Schematic of the continuous-time Markov process that is applied on a regular microscopic grid [fine grid lines]. The heavy grid lines represent a climate model-gridbox. b) The process describes area fractions of clear sky sites [0], cumulus congestus clouds [1] deep cumulus clouds [2] and stratiform clouds [3]. c) the calculation of the transition rates between the different types is based on a Markov chain that determines the allowed transitions between the types.}
\label{fig:5-06}
\end{figure}

\citet{Khouider2010} followed a more holistic approach and presented a stochastic process to represent the occurrence of different cloud types. Based on observations of the cloud characteristics in the tropics \citep{Johnson1999}  the SMCM applies a continuous-time Markov chain that describes the evolution of three different cloud types on a regular \emph{microscopic} grid of $N$ cells within a climate model grid box. Each cell can either be in \emph{clear sky} [0] or occupied by a \emph{congestus} cumulus [1], \emph{deep} cumulus cloud [2] or a \emph{stratiform} anvil [3, see also Figure \ref{fig:5-06}]. 

The model calculates transition rates $R_{kl}$ from cloud type $k$ to cloud type $l$. These rates are functions of the state $X$ [clear sky, congestus, deep, stratiform] of the stochastic process and an external potential $U$ that describes the large-scale state of the atmosphere [$R_{kl} = R_{kl}(X,k,l,U)$]. The Markov property implies that the transition rate of state $k$ to any other state is balanced by the transition rate from any other state to $k$ \citep{Pollett1987}. Figure \ref{fig:5-06}-c illustrates the transition rules and identifies transitions that are assumed not to occur, which are:
\begin{linenomath*}
\begin{equation*}
R_{03} = R_{21} = R_{32} = R_{31} = R_{13} = 0.
\end{equation*}
\end{linenomath*}
Following \cite{Khouider2014} the cloud area fraction of one of the three cloud types is equal to the conditional expected value [$\rho_i$] of all transition rates associated with each cloud type:
\begin{linenomath*}
\begin{equation}
\overrightarrow{CAF} = \begin{pmatrix}
\rho_1 \\ \rho_2 \\ \rho_3 
\end{pmatrix} = 
\dfrac{1}{N}\begin{pmatrix}
\frac{R_{01}}{R_{12}+R_{10}} \\
\Bigl(R_{02}+\frac{R_{01}R_{12}}{R_{12}+R_{10}}\Bigl)\cdot \Bigl(R_{23}+R_{20}\Bigl)^{-1} \\
\frac{R_{23}}{R_{30}} \Bigl( R_{02}+\frac{R_{01}R_{12}}{R_{12}+R_{10}}\Bigl)\cdot \Bigl(R_{23}+R_{20}\Bigl)^{-1} 
\end{pmatrix}
\end{equation}
\end{linenomath*}
\cite{Khouider2010} formulated simple expressions for the transitions:
\begin{linenomath*}
\begin{equation}
\begin{matrix*}[l]
R_{01} =& {\tau_{01}}^{-1} \Gamma(C)\cdot \Gamma(D) & R_{02} =&  {\tau_{02}}^{-1} \Gamma(C)\cdot \Bigl(1-\Gamma(D)\Bigl)\\
R_{12} =&  {\tau_{12}}^{-1} \Gamma(C)\cdot \Bigl(1-\Gamma(D)\Bigl) & R_{10} =& {\tau_{10}}^{-1} \cdot \Bigl(1-\Gamma(D)\Bigl)\\ 
R_{20} =& {\tau_{20}}^{-1} \cdot \Bigl(1 - \Gamma(C)\Bigl) & R_{23} =& {\tau_{23}}^{-1} \ \ \ \ \ \ R_{30} = {\tau_{30}}^{-1} \\
\end{matrix*}
\label{Eqn:5-10}
\end{equation}
\end{linenomath*}   
$C$ and $D$ are parameters for atmospheric instability and dryness, both take values between $[0,2]$ and represents large-scale averages over the climate model grid-box. $\Gamma$ serves as an internal response function and has the following form:
\begin{equation}
\Gamma(x) = \max(1-e^{-x},0)
\end{equation}
\citet{Peters2013} and \citet{DeLaChevrotiere2016} showed that the variability of simulated tropical convection is significantly improved when the SMCM is tuned with observational data.
Because of its simplicity the method described above can without complications be adopted to represent the behavior of coastal tropical convection, which will be discussed next.  
\subsection{New Transition Rates for Coastal Convection}
 \citet{Bergemann2016} found that rainfall is less dependent on mid-tropospheric humidity when coastal processes are present. They hypothesized that this behavior can be explained by meso-scale land-sea interactions, like sea-breeze circulations, that tend to moisten the atmosphere on scales smaller than the typical resolution of a climate model grid box. The above introduced SMCM simulates three different cloud types and calculates birth-death and transition rates between them. A possible approach to make the SMCM suitable to simulate coastal convection is to increase or decrease the rates of the relevant transitions according to the strength of the local land-sea interactions. 

The first step is to apply the trigger function, derived in Section \ref{sec:trigger}, to decide on sea-breeze conditions. The second step then is to modify the transition rates by applying an additional function that changes the probability of any transitions if coastal effects are deemed important.

The starting point of the derivation for the new transition rates of \emph{coastal} convection is the \emph{original} version of the SMCM introduced by \citet{Khouider2010}. Taking previous studies about the organization of coastal convection into account \citep[][a.o.]{Simpson1980,Simpson1993,Qian2008,Hill2010} an adoption of the given transition rates that mimic coastal convection should take the following considerations into account:
 \begin{enumerate}
\item During a strong sea-breeze day clouds are organized along the associated sea-breeze convergence.\label{Enum:5-1}
\item In this rather small line of convergence clouds will grow deeper and more quickly either because of merging, humidity advection or the sea-breeze convergence itself.\label{Enum:5-2}
\item The deep clouds remain organized along the associated convergence line and the lifetime of the associated cloud ensemble  is increased.
\label{Enum:5-3}
\end{enumerate}  

Within the framework of the SMCM consideration \ref{Enum:5-1} implies that congestus clouds are more likely to be "born" when sea-breezes are active. Consideration \ref{Enum:5-2} implies that the transition rates from congestus to deep and the birth-rates of deep clouds are also increased. The dissipation rate of congestus and deep clouds should at the same time be decreased [\ref{Enum:5-3}]. We choose not to alter the transition rates for stratiform clouds, as those are directly related to the presence of deep convective clouds, the existence of which our approach will enhance.

When coastal effects are strong the diurnal cycle of deep convective clouds over land is opposite to that over the adjacent ocean. In the morning, clouds are usually focused over the coastal ocean with clear sky conditions over coastal land. In the evening, clouds are usually building up over land with suppressed conditions over the adjacent ocean. To incorporate this behavior into the coastal trigger function the following equation can be applied:
\begin{linenomath*}
\begin{equation}
\tilde{f}^j(t) = f(t) \cdot \Theta^{j} \cdot \frac{g_{\Delta\mathrm{T}}(t)}{{\Delta\mathrm{T}}} \cdot \frac{{|\vec{V}|}-g_{|\vec{V}|}(t)}{{|\vec{V}|}} 
\label{Eqn:5-11}
\end{equation}
\end{linenomath*}
With:
\begin{itemize}
\item $f(t) = $ the binary value of the trigger function [yes/no] that indicates days with strong coastal processes.
\item $g_{\Delta \mathrm{T}/|\vec{V}|}$ the large-scale conditions for thermal heating contrast $\Delta$T$(t)=\mathrm{T_{land}(t)-T_{ocean}(t)}$ and wind speed $|\vec{V}(t)|$.
\item $\Theta^{j}= \left\lbrace\begin{matrix}
1 & \ \mathrm{if\ \ }j\mathrm{\ is\ a\ land\ point \ at \ the \ coast}\\
-1 & \ \mathrm{if\ }j\mathrm{\ is\ an\ ocean\ point \ at \ the \ coast}
\end{matrix}\right.$
\end{itemize}

The combination of $g_{\Delta\mathrm{T}}(t)$ which is usually positive during the day and negative during the night and the matrix $\Theta^{j}$ should reproduce the spatial and temporal occurrence of convective clouds in coastal areas.  Because $\tilde{f}^j(t)$ can be positive or negative and the final transition probabilities have to be strictly positive an additional function is applied that maps $\tilde{f}^j(t)$ into a positive interval:
\begin{linenomath*}
\begin{equation}
\zeta(\tilde{f}^j) = \left[\left(\arctan[\tilde{f}^j+\tan(1)]+\frac{\pi}{2}\right)\cdot \frac{11}{9\pi}\right]^2
\label{Eqn:5-12}
\end{equation}
\end{linenomath*}
$\zeta$ is designed to become $1$ if coastal processes are weak or not existing and can be therefore integrated into the SMCM transition rates as multiplication factors. The reader is reminded that $\zeta$ serves as an additional function that is applied to alter the occurrence probability of clouds within the SMCM and \emph{not} as a predictor for cloud area fractions or coastal rainfall. The influence of the magnitude of the $\zeta$ on the occurrence will be investigated in detail in Section \ref{sub:5-Coastl_effects}.

Taking the  considerations \ref{Enum:5-1} to \ref{Enum:5-3} into account the new \emph{coastal} version for the transition rates become:
\begin{linenomath*}
\begin{equation}
\begin{matrix*}[l]
R_{01} =& {\tau_{01}}^{-1} \Gamma(C)\cdot \Gamma(D) \cdot \zeta(\tilde{f}^j)& R_{02} =&  {\tau_{02}}^{-1} \Gamma(C)\cdot \Bigl(1-\Gamma(D)\Bigl)\cdot \zeta(\tilde{f}^j)\\
R_{12} =&  {\tau_{12}}^{-1} \Gamma(C)\cdot \Bigl(1-\Gamma(D)\Bigl)\cdot \zeta(\tilde{f}^j) & R_{10} =& {\tau_{10}}^{-1} \cdot \Bigl(1-\Gamma(D)\Bigl) \cdot \zeta(-\tilde{f}^j) \\ 
R_{20} =& {\tau_{20}}^{-1} \cdot \Bigl(1 - \Gamma(C)\Bigl) \cdot \zeta(-\tilde{f}^j)& R_{23} =& {\tau_{23}}^{-1} \cdot  \Bigl(1 - {\Gamma}(C)\Bigl) \\
R_{30} =& {\tau_{30}}^{-1} \\
\end{matrix*}
\end{equation}
\end{linenomath*}
The multiplication of an additional factor $\zeta$ increases or decreases the presence of convective clouds when coastal effects are present as expressed by the coastal trigger function $\tilde{f}$. Yet the application of this equation system has one disadvantage. When the atmosphere is relatively dry and stable the associated $\Gamma$ functions become small and hence the product of the two would be even smaller. Since the product of the two $\Gamma$ functions is small a multiplication of the function for the coastal effects [$\zeta$] would have very little effect on transition rates. \citet{Bergemann2015} suggested that coastal rainfall and convection are affected by meso-scale moistening and destabilization from coastal effects. To address this issue an additional parameter is \emph{added} in the calculation of $\Gamma$:
\begin{linenomath*}
\begin{equation}
\tilde{\Gamma}(X,\tilde{f}^j) = \max(1-e^{-\max(X+\varepsilon\cdot\tilde{f}^j,0)},0).
\label{Eqn:5-13}
\end{equation}
\end{linenomath*}
$\varepsilon$ is a constant that increases or decreases the exponent by only a small increment. It can be considered as a parameter that describes meso-scale moistening and destabilization by coastal processes.
The final set of equations for the coastal version are compared to the  version of the SMCM by \citet{Khouider2010} in Table \ref{tab:5-03}.
\begin{table}
\caption{Comparison of the modified transition rates for coastal clouds [coastal] and the rates developed by \citet{Khouider2010} [original].}
\centering
\begin{tabular}{ r l l}\toprule
& coastal version & original version \\ \midrule
$R_{01}$ & $ \tau_{01}^{-1} \tilde{\Gamma}(C,\tilde{f}^j)\cdot \tilde{\Gamma}(D,-\tilde{f}^j)\cdot \zeta(\tilde{f}^j)$ &$ \tau_{01}^{-1} {\Gamma}(C)\cdot{\Gamma}(D)$ \\
$R_{02}$ & $\tau_{02}^{-1} \tilde{\Gamma}(C,\tilde{f}^j)\cdot \Bigl(1-\tilde{\Gamma}(D,-\tilde{f}^j)\Bigl)\cdot \zeta(\tilde{f}^j)$ & $\tau_{02}^{-1} {\Gamma}(C)\cdot \Bigl(1-{\Gamma}(D)\Bigl)$ \\
$R_{12}$ & $\tau_{12}^{-1} \tilde{\Gamma}(C,\tilde{f}^j)\cdot \Bigl(1-\tilde{\Gamma}(D,-\tilde{f}^j)\Bigl)\cdot \zeta(\tilde{f}^j)$ & $\tau_{12}^{-1} {\Gamma}(C)\cdot \Bigl(1-{\Gamma}(D)\Bigl)$ \\
$R_{23}$ & $\tau_{23}^{-1} \cdot \Bigl(1-{\Gamma}(D)\Bigl)$ & $\tau_{23}^{-1} \cdot \Bigl(1-{\Gamma}(D)\Bigl)$ \\
$R_{10}$ &  $\tau_{10}^{-1} \cdot \Bigl(1-\Gamma(D)\Bigl) \cdot \zeta(-\tilde{f}^j) $ & $ \tau_{10}^{-1} \cdot \Bigl(1-\Gamma(D)\Bigl)$ \\
$R_{20}$ & $ \tau_{20}^{-1} \cdot \Bigl(1 - {\Gamma}(C)\Bigl) \cdot \zeta(-\tilde{f}^j)$ & $\tau_{20}^{-1} \cdot \Bigl(1 - {\Gamma}(C)\Bigl)$ \\
$R_{30}$ & $\tau_{30}^{-1}$ & $\tau_{30}^{-1}$ \\ \bottomrule
\end{tabular}
\label{tab:5-03} 
\[\tilde{\Gamma}(X,\tilde{f}^j) = \max(1-e^{-\max(X+\varepsilon\cdot\tilde{f}^j,0)},0) \ \ \Gamma(X) = \max(1-e^{-X},0)\]
\[ \zeta(\tilde{f}^j) = \left[\left(\arctan[\tilde{f}^j+\tan(1)]+\frac{\pi}{2}\right)\cdot \frac{11}{9\pi}\right]^2\]
\end{table}
It is evident that the coastal version reduces to the original set of equations when coastal effects are deemed to be absent [$\zeta(\tilde{f}^j=0)=1$].

\section{Simulation of Tropical Coastal Convection}
\label{sec:results}
The overall goal of this study is to introduce of a possible modeling approach that represents the behavior of clouds and convection in coastal tropical areas. It has been shown that this behavior can vary strongly with the presence and strength of coastal effects like land-sea breeze circulation systems. To investigate the influence of the strength of the coastal effects, expressed by the magnitude of the coastal trigger function $\tilde{f}$ and the additive constant $\varepsilon$ in the stochastic cloud model introduced above, a sensitivity analysis of the two parameters is presented here first. The results of the simulation with the \emph{coastal} version of the SMCM, abbreviated with SMCM\emph{-C} will be contrasted with those obtained with the original version which is labeled as SMCM-\emph{O}.
\subsection{Model Parameters}
To test various aspects of coastal convection in the above described parametrization framework we conduct several experiments. Before discussing this experiments we briefly outline the setup and parameters which are the basis of this experiments.
\subsubsection{Atmospheric Forcing Data}
If not mentioned otherwise the model is forced with atmospheric data, taken from the ERA Interim reanalysis project [ERA-I]. Recall that the ERA-I data has a resolution of 0.75$^\circ$ in space and 6 hours in time. One shortcoming of using reanalysis data is that feedback between simulated cloud cover and the strength of the coastal effects cannot be taken into account. This can only be done when the presented method is coupled to a GCM setup. We would like to remind the reader that the presented method should serve as one possible outline of how to tackle the problems associated with convection in the coastal tropics. It is not the aim of this study to present an entire parametrization for coastal convection in a GCM.

The SMCM considers tropospheric dryness and instability. The former can be expressed by the vertically integrated saturation fraction $r = \int q dp\ / \int q_s dp$. Here $q$ and $q_s$ are specific humidity and saturation specific humidity on pressure level [$p$]. $r$ has been chosen as a proxy for atmospheric moisture because \citet{Bretherton2004} showed an empirical exponential relationship between the variable and precipitation in the tropics. Guided by the work of \citet{Tan2013}, the modified k-index $ki = \frac{1}{2}\cdot(T_{sfc} + T_{850}) - T_{500} + \frac{1}{2} \cdot (T_{D_{sfc}} + T_{D_{850}} ) - (T_{700} - T_{D_{700}})$ is used to describe the atmospheric instability. Here $T_p$ and $T_{D_{p}}$ are the air and dew point temperatures at pressure level $p$. The k-index \citep{Cherba1977,Peppler1988} is frequently used in assessing the potential for the existence of thunderstorms in weather forecasting. It is chosen over the more commonly used convective potential energy [CAPE] as it is simpler to calculate but still achieves forecast skills of thunderstorm activity that are comparable with CAPE \citep{Haklander2003}. $r$ and k-index haven't been applied as forcing variables in the SMCM before, yet we only wish to investigate the impact of the coastal trigger function. Exploring the effects of the $r$ and k-index as large-scale atmospheric proxies is out of scope of this study.

In both versions of the SMCM the dryness and instability are dimensionless parameters that take values between 0 and 2. When driving the model with ERA-I data, the scaling of the two parameters becomes important. The dryness factor $D$ can be most simply calculated from the vertically integrated saturation fraction  $r$ given by:
\begin{linenomath*}
\begin{equation}
D = 2 \cdot (1 - r)
\end{equation} 
\end{linenomath*}
Unlike the saturation fraction, the k-index [$ki$] isn't bounded and has to be scaled to be between 0 and 2 as required by the SMCM. The scaling parameter is chosen to be 50\% of the 99th percentile of the global k-index climatology [$ki_{99}$]:
\begin{linenomath*}
\begin{equation}
C = \left(\frac{ki_{99}}{2}\right)^{-1} \mathrm{K^{-1}} \cdot \max\left(ki,0\right) = \frac{1}{23.5} K^{-1} \cdot \max\left(ki,0\right).
\end{equation}
\end{linenomath*}

\subsubsection{Model Domain and Resolution}
The model domains in the test cases are chosen to be the size of a global climate model gridbox, $\approx$150 km. For simplicity the domain is divided by a straight vertical coastline into land and ocean. The atmospheric forcing data is calculated from domain averages of tropospheric dryness $D$, instability $C$ and thermal heating contrast $\Delta T$. The micro grid that applies the Markov chain contains 100 lattice points which corresponds, given the domain size of 150 km, to a resolution of 1.5 km. The internal timestep of the model is chosen to be 10 minutes. If not mentioned otherwise the 6 hourly atmospheric conditions from ERA-I are interpolated to match the 10 minute time stepping of the SMCM. Table \ref{tab:5-04} summarizes the parameters of the optimized configuration setup. 

\subsubsection{Convective Time-Scales}
The SMCM has several constants that have to be chosen to represent the behavior of tropical convection. One important set of parameters that have been discussed in previous studies are the transition time-scales $\tau_{kl}$. In the past the choice of the time-scales was based on ad hoc intuition \citep{Khouider2010}. \cite{Peters2013} were the first to investigate the influence of the time scales on the model's representation of convection. They systematically tuned the time-scales to match statistics of simulated cloud area fractions with observations. \citet{DeLaChevrotiere2014} have  developped a rigorous Bayesian inference technique to learn these parameters from data. Unfortunately, this method has been applied so far to only large-eddy simulation data and not yet to actual observations.  In the present work the time-scales have been chosen according to \cite{Peters2013} and are given in Table \ref{tab:5-04}. 
\begin{table}
\caption{The fixed model parameters of time stepping $\Delta t$, domain size $dZ$, number of micro-grid points/horizontal resolution for given domain size $N$/$\Delta x$ and the convective time scales $\tau_{kl}$ for each transition; with 0 being clear sky , 1 congestus, 2 deep and 3 stratiform clouds.}
\centering
\begin{tabular}{*{10}{c}}\toprule
$\Delta t$ & $dZ$ & $N/\Delta x$ &\multicolumn{7}{c}{Transitions time scales ($\mathbf{\tau}$)} \\ \midrule
 & & & $0\rightarrow 1$ & $0\rightarrow 2$ & $1\rightarrow 2$ & $2\rightarrow 3$ & $1\rightarrow 0$ & $2\rightarrow 0$ & $3\rightarrow 0$ \\
10 min & 150 km &100/1.5 km  & 1 h & 2.2 h & 1.2 h & 0.16 h & 1.2 h & 2.4 h& 4 h\\ \bottomrule
\end{tabular}
\label{tab:5-04}
\end{table}
\subsection{The Influence of Coastal Effects}
\label{sub:5-Coastl_effects}
As demonstrated by \cite{Bergemann2016} coastal rainfall and with it deep coastal convection occur in drier environments than its open ocean or land counterparts. We recall that in the SMCM-C, the impact of land-sea interactions is described by the trigger function $\tilde{f}$ for coastal effects and the added parameter $\varepsilon$. To test the influence of these two parameters the SMCM is driven with dryness and instability conditions using a wide range of values for $\varepsilon$ and $\tilde{f}$ and the results are compared to satellite based estimates \citep[CMORPH][]{Joyce2004}. CMORPH has a spatial resolution of 0.25\textdegree with a time resolution of 3 hours. For this sensitivity test the large-scale atmosphere should be relatively dry and stable while still producing a considerable amount of rainfall. The conditions in the Sarawak province on Borneo [3.5\textdegree{}N, 113.2\textdegree{}E] during September 2000 were found to meet those criteria. 

During the considered period the k-index and $r$ remain below the 30th percentile of the ERA-I climatology (1986-2016)  while the rainfall is above the 70th percentile. It has also been shown that strong coastal effects in this region contribute to the high precipitation amount that is received throughout the year \citep{Geotis1985,Ichikawa2006,Qian2013,Bergemann2015}.

The strength of the coastal effects $\tilde{f}$ [calculated by the trigger function introduced above] is also estimated from ERA-I observations. To vary its magnitude and investigate the influence on the simulation of coastal clouds, $\tilde{f}$ is multiplied by a factor that describes the strength of coastal effects in each sensitivity run:
\begin{equation}
\tilde{c}(t) = A \cdot \tilde{f}(t).
\label{Eqn:5-14}
\end{equation}
In the following sensitivity analysis, the three parameters $C$, $D$ and $\tilde{f}$ are calculated from ERA-I data whereas the magnitude of the coastal effects $A$ in Equation \ref{Eqn:5-14} and the constant $\varepsilon$ in Equation \ref{Eqn:5-13} vary from simulation to simulation.

To determine optimal magnitudes of the two newly introduced variables the time series of cloud area fractions over coastal land are correlated with CMORPH rainfall occurring over the land parts marked in the purple box over Borneo in Figure \ref{fig:5-04}.  We use rainfall data as it serves as a good surrogate for deep convection in the tropics \citep{Davies2013}.

\begin{figure}[t]
\centering
\includegraphics[width=\textwidth]{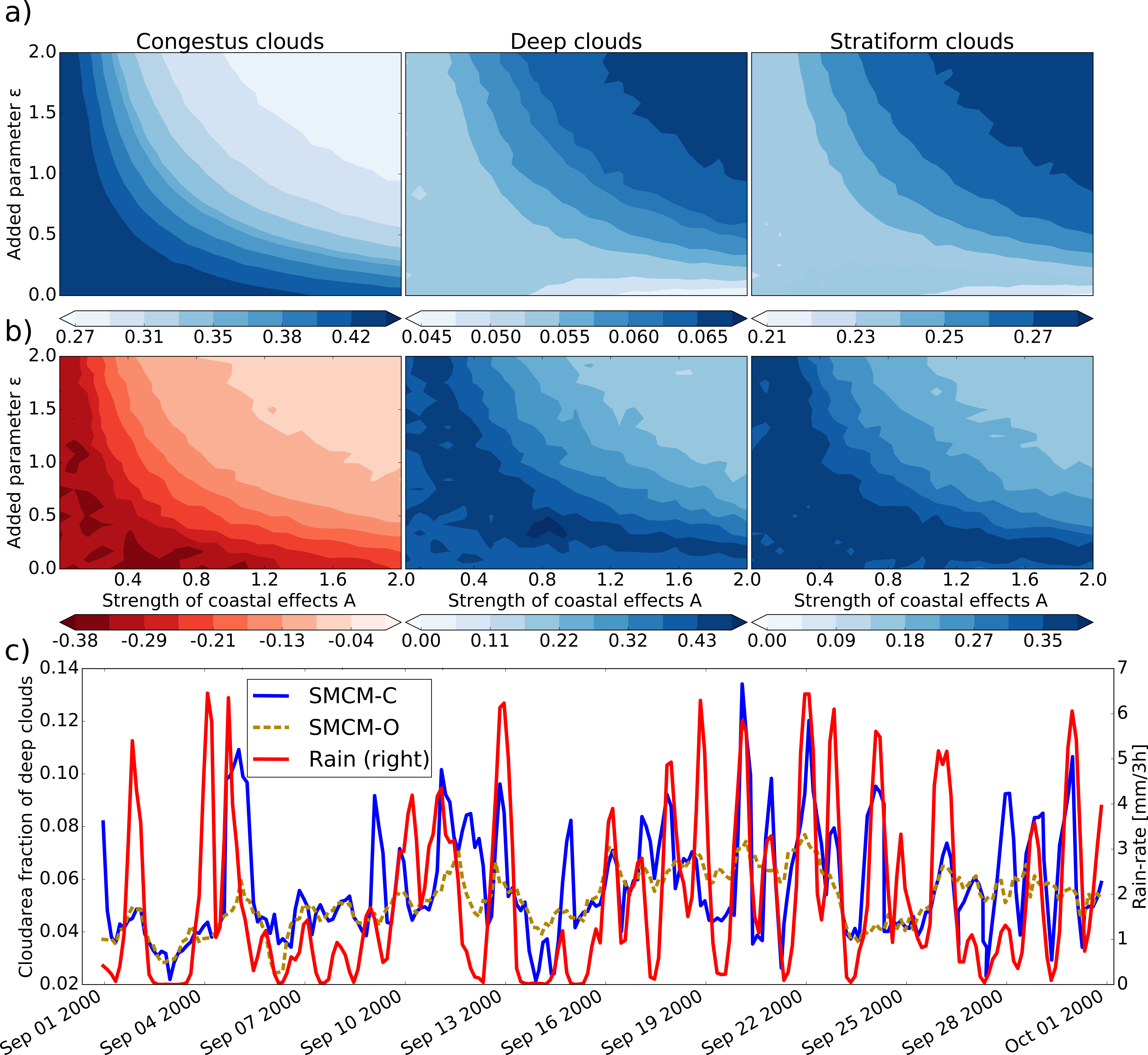}
\caption{a) Mean cloud area fraction as a function of amplitude of the coastal effects. b) The Pearson correlation coefficient of the time series of area averaged cloud fraction over coastal land and observed precipitation from satellite rainfall estimates over coastal land. c) Time series of precipitation [CMORPH satellite based rainfall estimated - see text for details, cloud area fraction of deep clouds over coastal land in the SMCM-O and SMCM-C [$A=0.8,\ \varepsilon=0.45$].}
\label{fig:5-08}
\end{figure}

The mean cloud area fraction for the simulated time period over Borneo from 1st to 30th of September 2000 as a function of the added constant $\varepsilon$ and the strength of the coastal effects $A$ are shown  in Figure \ref{fig:5-08}-a. Like in Section \ref{sec:trigger} the test domain is chosen to be half land and half ocean. The results for the SMCM-O are located at $A=\varepsilon=0$.
If both $\varepsilon$ and the magnitude of the coastal effects $A$ are weak the conditions are more favorable for congestus clouds while the area covered by deep and stratiform clouds is relatively small. The large-scale forcing during the simulated time period is relatively stable and dry, as a consequence the simulated cloud cover in the SMCM-O[($A=\varepsilon=0$] is mostly dominated by cumulus congestus clouds. An increase of both $\varepsilon$ and $A$ leads to higher birth rates of deep convective clouds and an increase of cloud area fraction of this cloud type. The growth of deep clouds is accompanied by an increase of area that is covered by stratiform clouds. If $A$ and $\varepsilon$ are chosen to be high, deep convective and stratiform clouds become the dominant cloud types and fewer congestus clouds are present. For low $\varepsilon$ values the build up and dissipation of clouds happens very rapidly when the amplitude of coastal effects is increased. Consequently there is a strong difference of cloud area fraction between land and adjacent ocean which leads to an overall decrease of mean cloudiness in the region. This behavior is indicated by a decrease of mean cloud area fraction of deep and stratiform clouds when $\varepsilon$ remains small and the strength of coastal effects is increasing.

We now decide on optimal magnitudes of the two newly introduced variables by correlating CMORPH precipitation with cloud area fractions of deep clouds. The correlation coefficient rather then standard verification methods like critical success index or root mean square error was chosen because the model predicts cloud area fractions and not precipitation. During the simulation period the rainfall observations were mainly over land and therefore only the conditions over coastal land rather than averages over land and ocean were taken into account.  Because congestus clouds peak about 3 to 5 hours earlier than the peak in rainfall and are relatively short lived they usually exhibit a minimum in area fraction when the peak in rainfall occurs. As a consequence the rainfall correlation is less than zero throughout the experiment. The correlations for deep and stratiform clouds have similar shapes with maximum rainfall correlations of up to $\approx$ 0.5 for deep convective clouds. The local correlation maximum that can be observed for deep clouds suggests an optimal range of the magnitudes for $\varepsilon$ and $A$. The overall maximum of correlation occurs at $A = 0.8$ and $\varepsilon= 0.45$. Decreasing the simulated time-period as well as varying the simulated location doesn't change these magnitudes significantly. The time series of observed precipitation and cloud area fraction of deep clouds in SMCM-O and SMCM-C is displayed in Figure \ref{fig:5-08}-c. Here, the optimal values of $A$ and the parameter $\varepsilon$ have been used [$A=0.8, \ \varepsilon= 0.45$]. The SMCM-C, unlike SMCM-O, has a much more distinct rainfall pattern that is relatively close to the observed rainfall. Most of the strong peaks in rainfall, up to 7 mm/3h, are accompanied by deep clouds in the SMCM-C. 

\subsection{The Relationship of Coastal Convection and the Large-Scale Atmosphere}
\begin{sidewaysfigure}
\centering
\includegraphics[height=0.5\textheight]{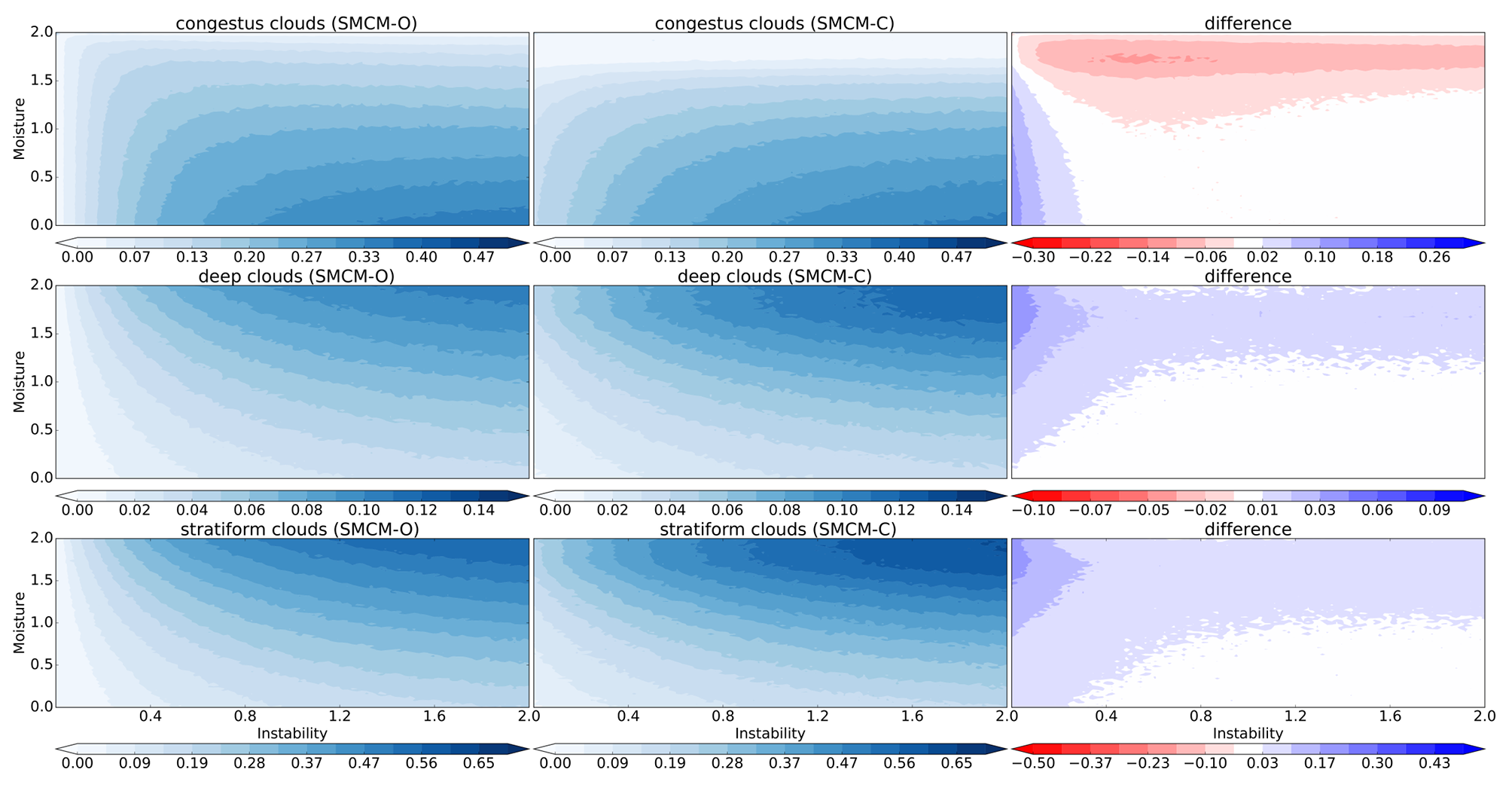}
\caption{Equilibrium cloud area fractions of congestus, deep and stratiform clouds as a function of moisture and instability.}
\label{fig:5-09}
\end{sidewaysfigure}
To test the influence of both, moisture and instability, the cloud area fractions of the three different cloud types in SMCM-O and SMCM-C are presented as a function of the two large-scale variables. Figure \ref{fig:5-09} displays the equilibrium cloud area fraction as a function of scaled moisture [1 - dryness] and instability of the three different cloud types that are predicted by the SMCM-O and SMCM-C with the optimal parameter setting for $A$ and $\varepsilon$ derived above.

If the conditions are drier and stabler [$\leq 1$] the SMCM-C produces more congestus clouds [upper panel in Figure \ref{fig:5-09}]. This leads to an increased transition rate of deep clouds in slightly drier and more stable atmospheres. When the large-scale environment becomes moister more deep clouds are directly 'born' from clear sky in the SMCM-C. The increased birth rate of deep cloud explains the decrease of the area that is covered by congestus clouds [upper right panel in Figure \ref{fig:5-09}]. The presence of deep clouds also leads to an increase of the transition to stratiform clouds.  With increasing instability and moisture both model versions show an increase of area that is covered by deep and stratiform cloud types. The difference between the two model versions decreases when the large-scale atmosphere becomes the dominant factor to produce deep and stratiform clouds. 

\citet{Bergemann2016} showed that when coastal effects are present convection is less dependent on the large-scale state of the atmosphere. To test whether the SMCM-C is able to capture this weaker atmospheric state to convection relationship we drive the model with ERA-I data from all tropical coastal locations. The simulated time period is January 1998 to December 2016. To combine and compare conditions across different geographical regions, cloud area fractions of deep clouds are grouped by their strength. This is done by calculating quintiles of cloud area fractions over the simulated domains and analyzing the distributions of instability, vertical velocity [$\omega$] and atmospheric humidity as a function of cloud area fraction quintile. 
\begin{figure}
\centering
\includegraphics[width=\textwidth]{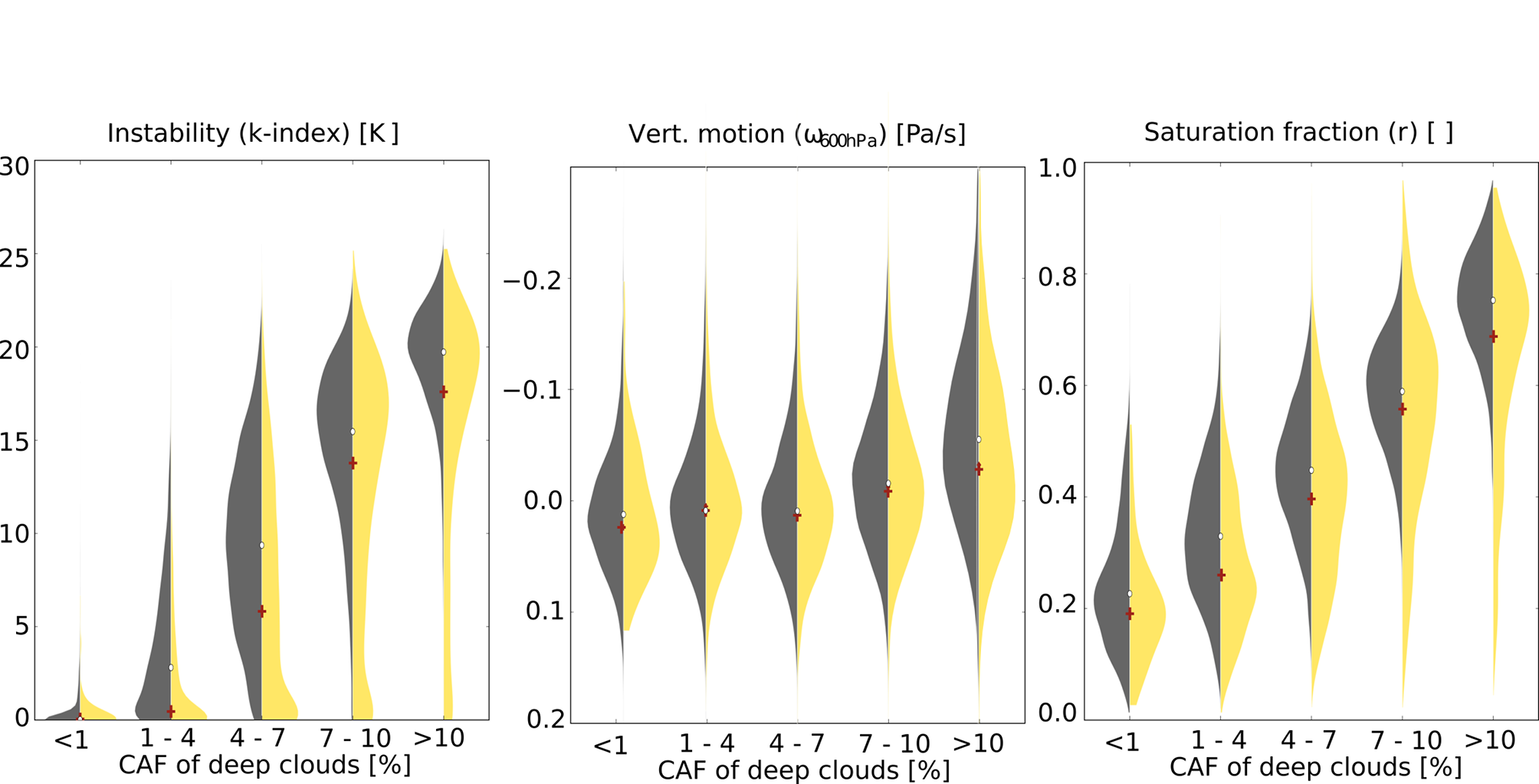}
\caption{Distributions of the the atmospheric variables $k-index$, $\omega$ at 600 hPa and $r$ within 5 quintiles of cloud area fractions [CAF] of deep convective clouds.  The Figure compares the distributions of the simulated cloud cover in the SMCM-O [black] and in the SMCM-C [yellow]. Medians are indicated by white dots for the left distributions and brown crosses for the right distributions.}
\label{fig:5-10}
\end{figure}
Figure \ref{fig:5-10} compares the distributions of instability represented by the modified  $k-index$ , large-scale vertical velocity $\omega$ at 600 hPa and atmospheric humidity expressed by the saturation fraction $r$, as a function of deep convective cloud area fractions in the SMCM-O [black] and in the SMCM-C [yellow]. Here, only events where the absolute value of the coastal trigger function differs from 0 and the convection should be supported by coastal effects are taken into account. Both distributions of the k-index and saturation fraction in the SMCM-O exhibit a similar behavior. When only few clouds are present the probability density functions [PDFs] in Figure \ref{fig:5-10} tend to be long tailed with relatively low medians of the atmospheric values. More clouds are associated with an increase in instability and moisture as indicated by the medians. With increasing cloud area fraction the distributions become more Gaussian. This is in contrast with the distributions of SMCM-C [yellow PDFs in Figure \ref{fig:5-10}]. Here the PDFs have longer tails and the increase of medians with increasing cloud-cover is weaker than in the SMCM-O version.  While less discernible, a signal similar to humidity and instability exists for vertical motion [center in Figure \ref{fig:5-10}]. The PDFs for the \emph{coastally} influenced clouds are shifted towards more subsidence/weaker ascent. The shape of all \emph{coastal} distributions and the comparisons of the medians for both SMCM-C and SMCM-O suggests that atmospheric instability, vertical motion and humidity are distinctly different when the convection is likely to be supported by coastal processes and hence modeled deep convection can occur in more stable and drier atmospheres that have only weakly large-scale ascending motion.

The analysis shows that the modified SMCM-C is able to increase the occurrence of deep clouds in drier and more stable large-scale environmental conditions once land-sea interactions are present. As it has been shown in \citet{Bergemann2016} this behavior is one of the important differences of convection in the coastal tropics to that over the open ocean and inland areas.

\subsection{The Diurnal Cycle}
\label{Diurnal_cycle}
The above analysis shows that the SMCM-C can capture some important features of coastal convection in the tropics. We now turn our attention to the simulation of the diurnal behavior of convection in the SMCM-C. The diurnal cycle of convection, especially that of deep convective clouds, is one of the most prominent features of coastal convection. When coastal effects are strong clouds usually form over the ocean during night and early morning. During the day convection is mainly focused over land, peaking in the late afternoon to evening. Any modeling approach that is designed to capture the characteristics of coastal convection should be able to represent this spatial and temporal behavior. 

\begin{figure}
\centering
\includegraphics[width=\textwidth]{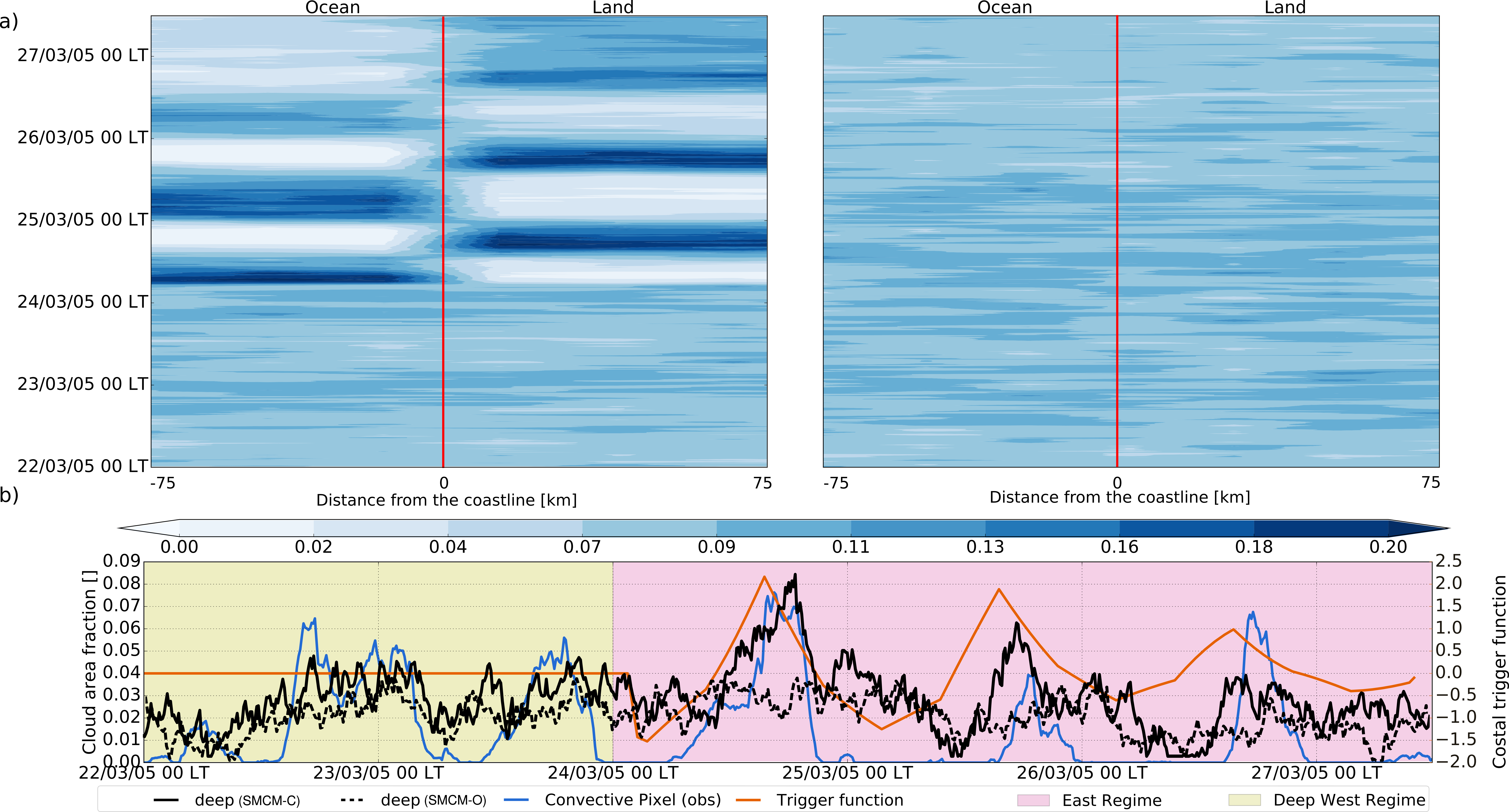}
\caption{a) The diurnal cycle of the mean cloud area fraction along the coastal cross Section the left hand side represents the SMCM-C while the right hand side shows the SMCM-O. b) Domain average of deep convective clouds over land [black], the value of the trigger function for coastal effects [orange], the CPOL dual band radar observations of area fraction of convective pixels over land [light blue].}
\label{fig:5-11}
\end{figure}

To study the diurnal cycle of clouds both versions of the SMCM are driven with atmospheric conditions, derived from ERA-I, over Darwin, Australia [12.5\textdegree S,130.9\textdegree E]. The test period is chosen to be the end of the wet season where the conditions often alternate between those strongly influenced by meso-scale land-sea interactions and those dominated by the large-scale environment \citep{Keenan2008,Pope2009}. For simplicity the topography is assumed to be a straight coastline that divides the domain into ocean and land. Figure \ref{fig:5-11}-a shows a cross Section of the simulated cloud area fraction of deep clouds.

The beginning of the simulation period is characterized by easterly winds advecting moisture from the Maritime Continent. These moist and unstable large-scale conditions are associated with high cloud area fractions of deep clouds in both versions of the SMCM. The cloud cover prevents the development of a strong thermal heating contrast between land and ocean, consequently coastal effects are weak and the trigger function is zero. In the absence of land-sea interaction the simulated cloud area fractions of both model versions do not differ. On the 23rd of March the wind regime shifts and slightly drier air is advected into the region. The associated decrease of large-scale humidity and instability leads to a decrease of area that is covered by deep clouds. The thermal heating contrast increases and coastal effects become stronger. Now the SMCM-C exhibits a very dominant diurnal cycle with deep clouds occurring over land  during the day and a moderate offshore propagation with peaks over the adjacent ocean in the early morning [Figure \ref{fig:5-11}-b]. This is in stark contrast to the SMCM-O where the number of deep convective clouds shows neither a spatial nor a temporal variation until the large-scale forcing increases slowly, which is accompanied by an increase of the number of deep clouds. Comparing the simulated cloud area fractions over land with observations of convective pixels over land from a dual band radar that operates across the simulated domain shows that the strong diurnal cycle that is associated with the east regime is better captured by the SMCM-C. 

Although some cloud propagation can be seen between land and ocean when coastal effects are present the main convection characteristics is dominated by a rather binary on and off behavior over land and ocean [Figure \ref{fig:5-11}-a]. This causes a rather artificial minimum of convection in the transition phase from land to ocean when the model has little clouds over the ocean while it hasn't sufficient clouds over land yet. Taking the interaction between neighboring grid cells into account can potentially improve this behavior \citep[see][ for details on local interaction]{Khouider2014}. Yet the influence of local interaction on the accuracy of the representation of convection in the SMCM hasn't been studied and is beyond the scope of this very first investigation.  Despite these issues and despite the relatively coarse 6 hourly resolution of the ERA-I input data the SMCM-C reasonably simulates the diurnal cycle of deep convection when coastal effects are strong in the Maritime Continent region \citep[e.g][]{Mori2004,Rauniyar2010}.
\subsection{The Influence of Randomness}
In the SMCM random numbers play an important role to determine the transition time [when does a cloud transition occur?], the micro-site [where on the microscopic grid does the transition occur?] and the transition type. To test the influence of stochasticity on the model results we create an ensemble of 30 members and vary the seeds of the generator that creates the pseudo random numbers.   Usually a random number generator applies some kind of a periodic sequence with a given seed number as initial value \citep[e.g][]{Wolfram1983,Matsumoto1998}. Fixing this seed makes the random number generator by definition deterministic and can thus have profound impact on the simulation. We chose to increase the seed numbers from 0 to 87 in increments of 3. The model setup is identical to the Darwin test case described above in Section \ref{Diurnal_cycle}. The reference simulation was carried out without fixation of the random number seeds. Here the seed numbers are automatically generated from various operating system events [gathered by \emph{/dev/urandom}].  

\begin{figure}
\includegraphics[width=\textwidth]{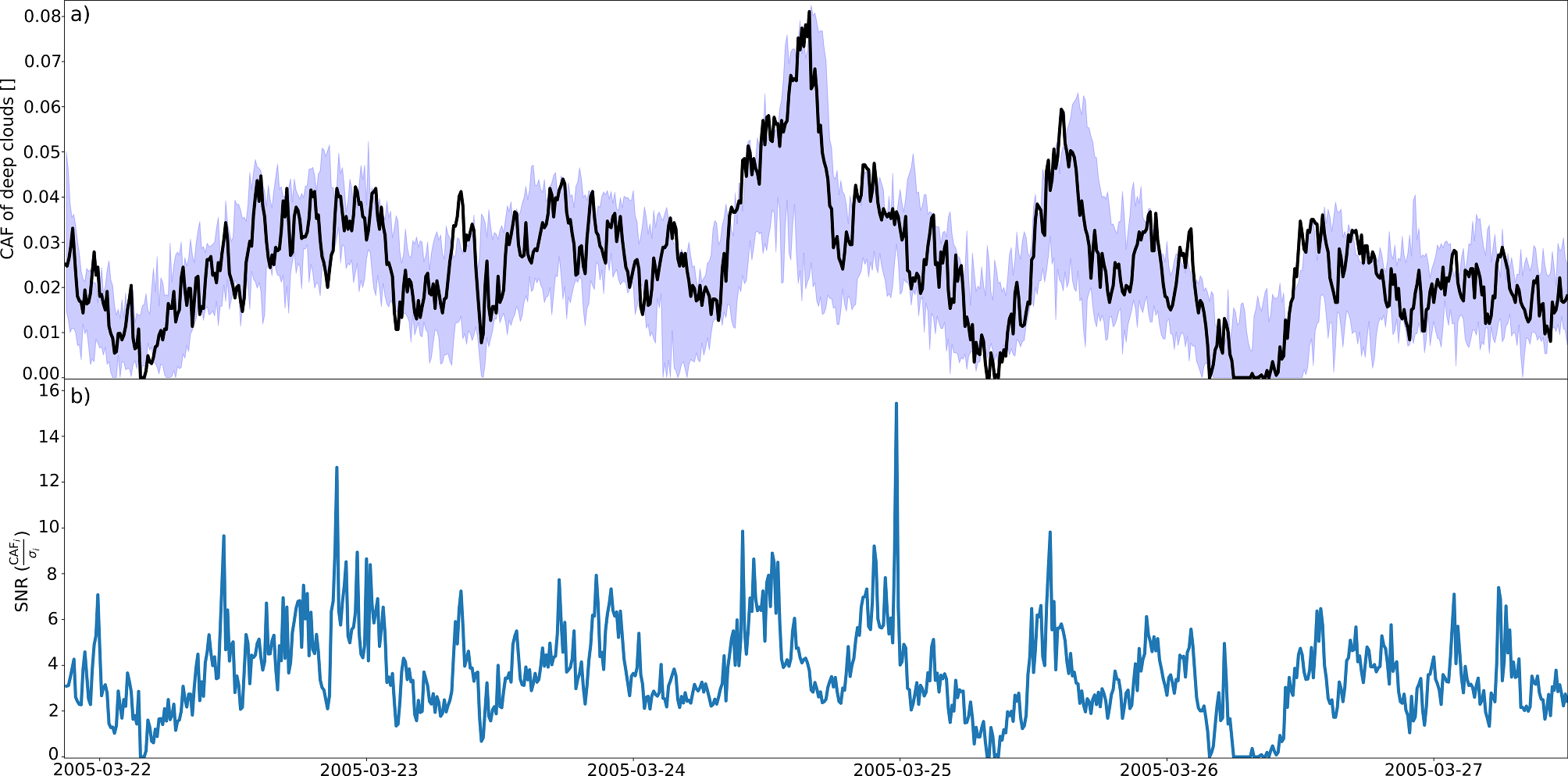}
\caption{a) Simulated area area fractions of deep clouds for the Darwin test case [see also Section \ref{Diurnal_cycle}]. The black line shows the reference run while the shading represents the ensemble that is created to test the influence of randomness in the model. b) Signal to Noise Ratio [SNR] defined as the cloud area fraction of deep clouds at timestep $i$ ($\mathrm{CAF}_i$) divided by the ensemble standard deviation at timestep $i$ ($\sigma_i$).}
\label{fig:5-09_A}
\end{figure}

The reference simulation [black line in Figure \ref{fig:5-09_A}-a] should not be considered as the  mean of the ensemble [blue shaded area in Figure \ref{fig:5-09_A}-a] because the nature of how random numbers are generated differ fundamentally. Therefore the reference simulation [with random seeds] lies for some occasions outside the ensemble [with fixed random seeds].Although there is a discernible spread among the ensemble members the overall signal is captured by all members. This is also shown by the signal to noise ratio [SNR] in Figure \ref{fig:5-09_A}-b. Here we define SNR by the area fraction of simulated deep clouds, $\mathrm{CAF}_i$, at timestep $i$ in the reference run divided by the ensemble standard deviation, $\sigma_i$, at timestep $i$. On average the forecasted cloud area fractions are $3.60\pm 1.90$ times higher than the ensemble spread at a given timestep. The intrinsic stochcasticity of the SMCM has only limited influence on the simulation of cloud area fractions in the model. Yet studying the influence of stochasticity when the SMCM is coupled to a state of the art GCM should be considered in future investigations.

\section{Summary and Conclusion}
\label{sec:con}
The aim of this study is to propose an idealized modeling approach and outline a parametrization framework that is able to represent coastal tropical convection in the context of global weather forecasting and climate models. Simulating coastal convection is a challenging task because it is strongly influenced by unresolved meso-scales. Additionally coastal effects are not always present. To address theses issues a trigger function that only depends on large-scale pre-cursors and identifies potential sea-breeze days was presented.

A sensitivity analysis has shown that thermal heating contrast and large-scale wind speed have most impact on the results when the trigger function is applied to atmospheric conditions in the tropics. The function was then scaled by the magnitude of wind speed and thermal heating contrast to get continuous rather than binary yes/no values. 

This continuous function has been applied in the stochastic multi cloud model that describes the dynamics of three different cloud types in the tropics: cumulus \emph{congestus}, \emph{deep} cumulus and \emph{stratiform} anvils. In its original form the model calculates birth, death and transition rates of the three cloud types based on the large-scale environmental conditions.  The transition rates were modified to mimic coastally influenced clouds by multiplying the continuous trigger function by the strength of the coastal effects. When the large-scale forcing is weak but coastal effects are strong a multiplication of an additional factor would be insufficient. An additional additive parameter describing meso-scale moistening and destabilization by land-sea interactions was introduced to change the large-scale conditions by only a small increment and increase the likelihood of convection in drier and more stable large-scale atmospheres. 

Applying the models to observations over Borneo showed that the \emph{coastal} model version [SMCM-C] simulates more clouds than the \emph{original} version [SMCM-O] when coastal effects are strong. A comparison with rainfall observations over a one month test period indicated that the model can improve the simulated occurrence of deep convective clouds. Testing the influence of large-scale atmospheric conditions on the dynamics of deep convective clouds suggested that the SMCM-C is able to capture the relationship of the large-scale atmospheric environment and coastally associated rainfall that has been identified in observations \citep{Birch2016,Bergemann2016}. 

Analyzing the diurnal cycle showed that the model is capable of representing the spatial and temporal behavior that is well known for clouds in the coastal tropics. It was also shown that when the convection is \emph{not} influenced by land-sea interactions the model shows no difference to the SMCM-O. This is an important step forward in the simulation, by a simple stochastic model, of the behavior of clouds in coastal areas, yet several challenges remain. 

The model does not correctly represent the propagation of clouds from land to the adjacent ocean and vice versa. This leads to a rather artificial decrease of clouds for times when there is little thermal heating contrast. This behavior could be improved by changing the influence of the neighboring clouds on convection \citep{Khouider2014}. The influence of nearest neighbor interaction hasn't been studied extensively and should be subject to future studies. 
\begin{figure}
\centering
\includegraphics[width=0.9\textwidth]{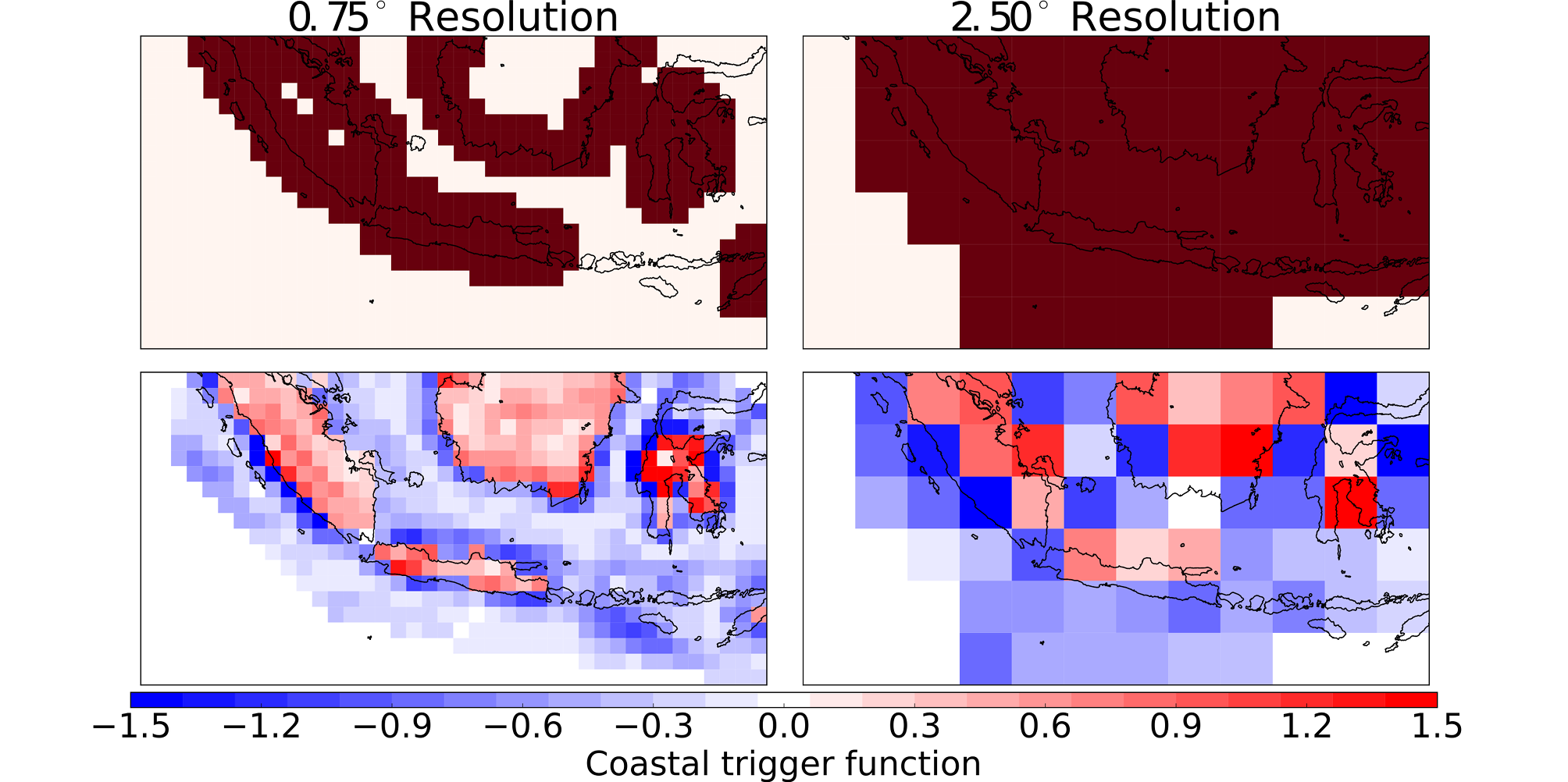}
\caption{A mask that defines coastal points and the strength of land-sea interactions by the magnitude of the coastal trigger function on the defined coastal grid points.}
\label{fig:5-12}
\end{figure}
Coastally associated rainfall is not only influenced by the presence of land-sea breeze circulations. Especially the orography can have considerable impact on the formation of coastal rainfall. Because the SMCM operates on microscopic lattice with a resolution of few kilometers the influence of orographic effects in coastal areas can in principle be added to alter the transition rates of clouds in coastal areas. Once again, this is a topic for future research. 

A question that remains unanswered is how to implement the adopted model version into a global model to improve simulations in coastal areas. The simplest approach would be defining an area that is affected by coastal effects and calculating the coastal trigger function in the associated climate model grid boxes. Studies have shown that coastal effects can occur roughly 150 to 250 km on- and offshore \citep[e.g][]{Mori2004,Keenan2008,Hill2010}. This area could be defined as the region where coastal effects are calculated in the model [Figure \ref{fig:5-12}]. Spatial interpolation of the coastal trigger function can be applied to increase the area that is affected by coastal effects and smooth the values of the coastal trigger functions towards inland and open ocean areas. The application of the trigger function is independent of the stochastic cloud model, and hence can also be implemented in existing cumulus parametrization schemes. For example it could be applied to increase the buoyancy of lifted parcels in mass flux based parametrization schemes. The stochastic multi cloud model itself has already been successfully applied in the global GCM ECHAM-6 by \cite{Peters2017}. Here the model serves as a closure term for cloud base mass-flux. \citet{Goswami2017} followed a different approach and  derived the convective heating rates in the Climate Forecast
System version 2 from cloud area fractions calculated by the SMCM.

 Because the presented \emph{coastal} version of the stochastic model differs only by a multiplicative factor it can without complications be extended to the existing model that has already been coupled to a GCM.
While our results are encouraging we note that the framework of an idealized model setup is only a first step intended to guide future work on how a possible parametrization of coastal convection and clouds could be designed. The results and the simplicity of the SMCM-C shows that the stochastic multi cloud model offers potential for further modification to parametrize and represent clouds and convection in coastal tropical areas.


%
%
%
%
\appendix
\section{Variance Based Sensitivity Analysis}
\label{sec:appendix}
A sensitivity analysis should give an estimate of the sensitivity of the output of a model with respect to changes in the input variables.  Giving an estimate of the sensitivity of a model can be useful for data assimilation, model tuning,  calibration, and dimensionality reduction. 

Formally the output $Y$ can be seen as the direct result of the applied model, in the present case the trigger function for coastal effects $f(\vec{X})$ , that takes the input $\vec{X}$ with $d$ uncertain input values $\vec{X}=\{X_1,..,X_d\}$. $f(\vec{X})$ can be decomposed by the following orthogonal functions:
\begin{linenomath*}
 \begin{equation}
 f(\vec{X}) = f_0 + \sum_{i=1}^d f_i(X_i) + \sum_{i<j}^d f_{ij}(X_i,X_j)+ \dots + f_{1,\dots,d}(X_1,\dots,X_d)
 \label{Eqn:5-2}
\end{equation}
\end{linenomath*}   
This decomposition states that the total output of a model $f(\vec{X})$ can be written as the sum of terms  measuring the dependence of the all independent input variables $X_i$ and the conditional changes of all input variables $(X_1,\dots,X_d)$.
If the values of $\vec{X}$ are independently and uniformly distributed on a unit hypercube $X_i \in [0,1] \forall i = 1,2,\dots,d$ then terms of higher order in Equation \ref{Eqn:5-2} vanish and the decomposition becomes orthogonal. The functional decomposition can be defined as:
\begin{linenomath*}
\begin{equation}
\begin{split}
f_0&= E(Y) \\
f_i(X_i)&= E(Y|X_i) - f_i \\
f_{ij}(X_i,X_j) &= E(Y|X_i,X_j) - f_0 - f_i - f_j
\end{split}
\label{Eqn:5-3}
\end{equation}
\end{linenomath*} 
Therefore the terms $f_i$ describe effects from the variation of the input value $X_j$ while $f_{ij}$ describes the effects of the variation of $X_i$ and $X_j$ at the same time. Square integration of Equation \ref{Eqn:5-2} gives:
\begin{linenomath*}
\begin{equation}
\begin{split}
\int_0^1 f^2(\vec{X})d\vec{X} - f_0^2 &= \sum_{i=1}^d \int_0^1 f^2_i(X_i) + \sum_{i<j}^d \int_0^1 f^2_{ij}(X_i,X_j) \\
\mathrm{Var}(Y) &= \sum_{i=1}^d V_i + \sum_{i<i}^d V_{ij}
\end{split}
\label{Eqn:5-4}
\end{equation}
\end{linenomath*}
The $V_i$'s can be seen as the variances due to variation of input variable $X_i$. Therefore the term: $S_i = \frac{V_i}{\mathrm{Var}(Y)}$ is the contribution of the variance of input $X_i$ to the total variance Var$(Y)$. It is important that the random input variables are uniformly and independently distributed in the $d$-dimensional input variable hypercube. This constraint is usually guaranteed by constructing a Sobol-sequence of length $N$ with the $d$ random input variables \citep{Sobol1967}.

\begin{figure}[h]
\centering
\includegraphics[width=\textwidth]{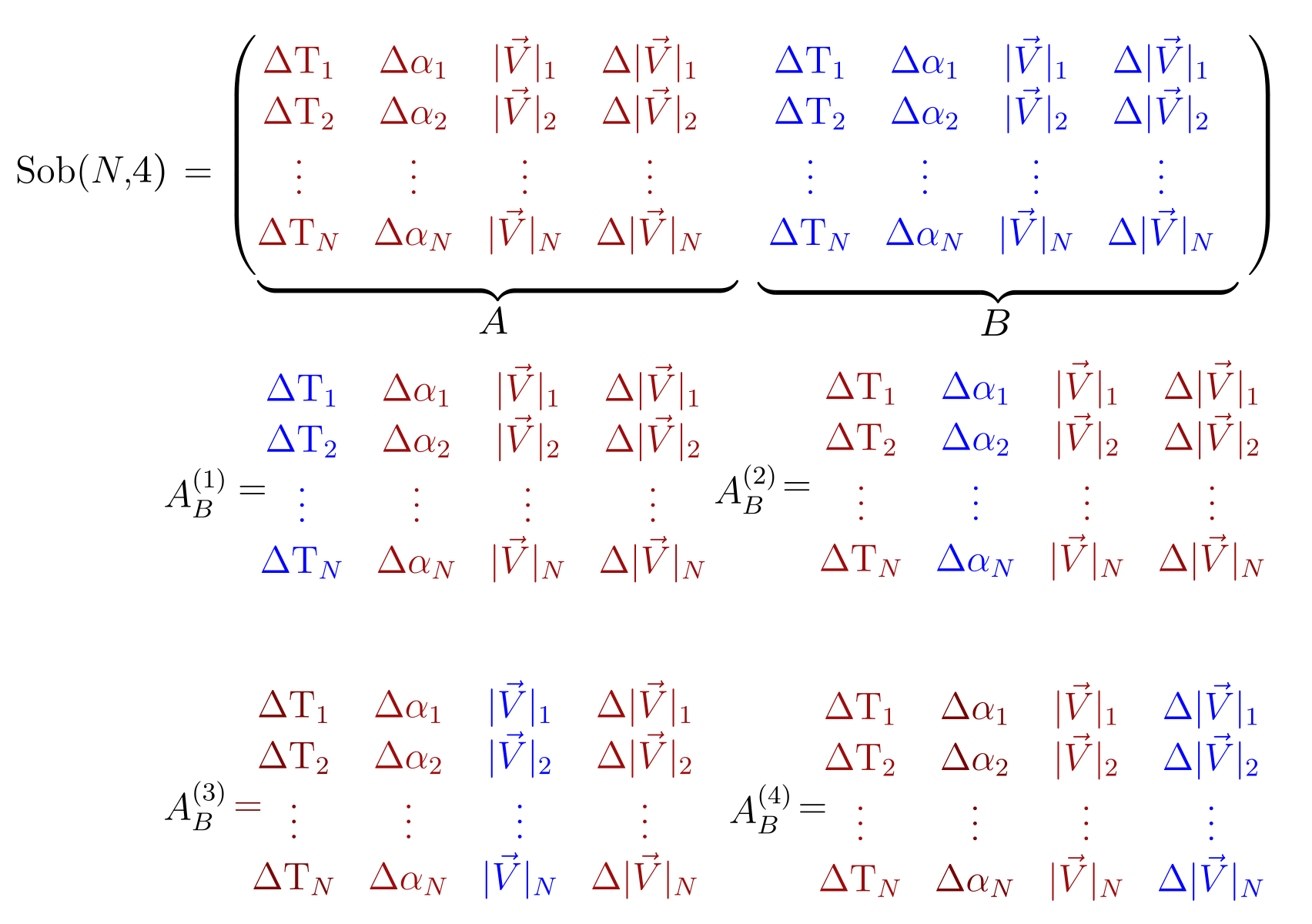}
\caption{Illustration of the construction of the $4$ $A_B^{(i)}$ input matrices from two $N\times 4$ dimensional input data matrices $A$ and $B$. The elements of the matrices are: $\Delta\mathrm{T}$ thermal heating contrast, $\Delta \alpha$ the change in large-scale wind direction, $|\vec{V}|$ the large-scale wind speed and $\Delta|\vec{V}|$ the change of large-scale wind speed.}
\label{fig:A-01}
\end{figure}
The following quasi-Mote-Carlo method is applied to calculate the values $V_i$ and $V_{ji}$:
\begin{enumerate}
\item[$i$] Two Sobol-sequences [$A$ and $B$] with four input variables are constructed. The input variables are the change in large-scale wind direction $\Delta \alpha$, the large-scale wind speed $|\vec{V}|$ the change in large-scale wind speed $\Delta|\vec{V}|$ and the thermal heating contrast between land and ocean $\Delta\mathrm{T}$ [see also Figure \ref{fig:A-01}]
\item[$ii$] 4 $N \times 4$ matrices are build from periodic permutations of the rows in $A$ and $B$ [see Figure \ref{fig:A-01}]
\item[$iii$] Run the filtering process $f$ with each of the input sets [rows] $A$, $B$ and the $d$ $A_B^{(i)}$.
\item[$iv$] Calculate the sensitivity indices with the following relation \citep{Saltelli2010}:
\begin{linenomath*}
\begin{equation}
S_i = \dfrac{V_i}{\mathrm{Var}(f(A))} \approx \dfrac{1}{\mathrm{Var}(f(A))}\dfrac{1}{N} \sum_{j=1}^N f(B)_j \cdot (f(A_B^{(i)})_j - f(A)_j)
\end{equation}
\end{linenomath*}
\end{enumerate}

\acknowledgments
We would like to acknowledge Todd P. Lane for his valuable suggestions that helped to realize develop the presented method. We are also very grateful to the two anonymous reviewers that have helped to improve the quality of the submitted manuscript. This research was supported in part by the Monash University eResearch Centre and eSolutions-Research Support Services through the use of the high-memory capability on the Monash University Campus HPC Cluster. We also acknowledge the Australian Research Council's Centre of Excellence for Climate System Science (CE110001028) for funding this work. The CMORPH satellite based rainfall estimates were obtained from the Climate Prediction Center (CPC) of the National Oceanic and Atmosphere Administration (NOAA). The Era-interim reanalysis data is supplied by the European Center for Medium Weather Forecast (ECMWF). The source code and a documentation of the algorithm that detects coastline associated rainfall can be retrieved from Zenodo (\texttt{http://dx.doi.org/10.5281/zenodo.44405}) or via GitHub (\texttt{https://github.com/antarcticrainforest/PatternRecog})

{\clearpage}





\listofchanges

\end{document}